\documentclass[10pt,reqno]{article}

\usepackage{fullpage}

\usepackage{amsmath,amsfonts,amssymb,amsthm}
\usepackage{mathtools}
\usepackage{stmaryrd}

\usepackage{cite}

\usepackage[unicode=true,hypertexnames=false]{hyperref}

\usepackage[dvipsnames,table]{xcolor}
\usepackage{graphicx}

\usepackage{float}
\usepackage{booktabs}
\usepackage{caption}
\usepackage{subcaption}
\usepackage{adjustbox}
\usepackage{diagbox}
\usepackage{blkarray}

\usepackage[ruled,vlined]{algorithm2e}

\usepackage{tikz}
\usetikzlibrary{
  trees,
  positioning,
  decorations.pathreplacing,
  angles,
  quotes,
  matrix,
  arrows.meta,
  decorations.pathmorphing,
  decorations.markings
}
\usepackage{tikz-cd}

\usepackage{bbold}
\usepackage{centernot}
\usepackage{enumerate}
\usepackage{enumitem}

\usepackage{todonotes}
\usepackage{comment}
\usepackage{soul}

\usepackage{authblk}

\usepackage{mypack}

\newcounter{commcount}
\definecolor{purpleish}{rgb}{0.5, 0.0, 0.5}
\definecolor{pinkish}{rgb}{0.8, 0.0, 0.4}

\begin{document}

\title{Stabilizer-Restricted Antidistinguishability in
Quantum Computation with Magic States}
\date{}
\author{Selman Ipek}
\affil{{\small \textit{Institut f\"ur Theoretische Physik, Leibniz Universit\"at Hannover, 30167 Hannover, Germany}}}

  \maketitle  
  \vspace{-4em}
\begin{abstract}
An antidistinguishing measurement for a family of preparations excludes at least one preparation for every possible outcome, {constraining `classical' representations of these processes, known as ontological models.}
We develop this connection within quantum computing with magic states (QCM), a universal model of quantum computation.
Our constructions use adaptive Pauli measurements on tensor products of single-qubit non-stabilizer states, including families associated with the Clifford hierarchy.
These constructions yield stabilizer-restricted versions of the Pusey–Barrett–Rudolph theorem: any ontological model reproducing these measurement statistics and satisfying preparation independence must assign disjoint ontic supports to the preparations considered.
We demonstrate this obstruction to preparation independence in a concrete ontological model based on convex polytopes used for the classical simulation of QCM.
\end{abstract}

  \tableofcontents

\section{Introduction}

A central theme in the foundations of quantum theory has been to identify a precise notion of what is meant by `classical'.
{Throughout the development of the subject, a web of such notions has emerged,}
based on locality~\cite{bell1964einstein}, noncontextuality~\cite{kochen2011problem,bell1966problem}, its generalized variant~\cite{spekkens2005contextuality},
{or nonnegativity in quasiprobability representations~\cite{wigner1932quantum,ferrie2008frame}.}
{A common framework for studying these notions describes an underlying physical reality in terms of an \textit{ontic state}~\cite{harrigan2010einstein}.}
{Quantum theory can be described in operational terms~\cite{hardy2001quantum},}
i.e. by preparations, transformations, and measurements,
{while ontological models represent these operational primitives using probability distributions, transition kernels, and response functions defined on the ontic state space.}
Additional constraints on the mathematical form of the ontological model encode the relevant notion of classicality,
with attendant `no-go' results demonstrating incompatibility with quantum mechanics.

Pusey, Barrett, and Rudolph (PBR)~\cite{pusey2012reality} proved a fundamental `no-go' theorem on the nature of the quantum state.
Their argument is based on the principle of preparation independence,
which reflects the classical intuition that operationally independent preparations should be represented in the ontological model by statistically independent product distributions.
The PBR theorem places particular emphasis on the ontological representation of preparations: under preparation independence, distinct pure quantum states must be represented by distributions with disjoint supports.
Another key notion is that of antidistinguishability~\cite{caves2002conditions,leifer2014quantum}.
A quantum measurement is said to be weakly antidistinguishing for a set of states if every measurement outcome has zero probability for at least one preparation in the set~\cite{srikumar2026contextuality}.
The existence of such a measurement forces the ontological representations of the preparations to have vanishing common overlap.
This is precisely the operational ingredient needed to prove the PBR theorem.

{Complementary to these foundational developments has been an increased interest in the role of such nonclassical features in quantum computational advantage.}
Within the setting of measurement-based quantum computation~\cite{raussendorf2001one}, Bell nonlocality and contextuality have been identified as resources for increased computational power relative to classical counterparts~\cite{anders2009computational,hoban2011non,raussendorf2013contextuality,frembs2023hierarchies}.
Similarly, negativity in quasiprobability representations quantifies the cost of particular classical simulation schemes~\cite{veitch2012negative,pashayan2015estimating,howard2017application,raussendorf2020phase}.
In several settings, contextuality and negativity have been shown to be equivalent notions~\cite{howard2014contextuality,delfosse2017equivalence,schmid2022uniqueness,raussendorf2017contextuality,okay2024classical}.
By contrast, the computational implications of the PBR theorem and its underlying principle of preparation independence have received less attention; a related connection to classical simulation is developed in~\cite{karanjai2018contextuality}.

Ontological models also provide a natural framework for studying classical simulation of quantum circuits~\cite{montina2008exponential,karanjai2018contextuality,kleinmann2011memory}.
A prominent example is the discrete Wigner representation of Gross~\cite{gross2006hudson}.
Restricted to states with nonnegative Wigner functions and stabilizer operations, it provides a noncontextual ontological model for odd-dimensional qudit systems that includes the stabilizer fragment~\cite{delfosse2017equivalence} and facilitates a sampling-based classical simulation scheme~\cite{veitch2012negative}.
In particular, the Wigner representation respects tensor products, so that independently prepared states with nonnegative Wigner functions are represented by product distributions.
This gives a preparation-independent model of the corresponding operational fragment.
{For states exhibiting negativity, the same tensor-product property enables a multiplicative measure controlling the sampling cost~\cite{pashayan2015estimating}.}

For qubits, many of these appealing properties of the odd-qudit case do not persist, and understanding these distinctions has been the subject of continued investigation~\cite{delfosse2015wigner,bermejo2017contextuality,raussendorf2017contextuality,lillystone2019contextual,raussendorf2020phase,raussendorf2023role}.
An important development along these lines was made in~\cite{zurel2020hidden} with the introduction of a family of geometric structures called Pauli polytopes, also known as $\Lambda$ polytopes, that are related by duality to the well-known stabilizer polytopes~\cite{heinrich2019robustness}.
These objects furnish a concrete $\psi$-epistemic ontological model that can be used for the classical simulation of quantum computing with magic states (QCM)~\cite{magic}, a universal model of quantum computation.
As such, these ontological representations can be rigorously studied using polytope-theoretic methods,
potentially yielding insights into the complexity of classical simulation and the boundaries of quantum advantage.

\section{Results and outline}

\subsection{Summary of main results}

In this paper we construct families of antidistinguishing measurements using adaptive and nonadaptive multiqubit Pauli measurements applied to preparation families consisting of completely separable states of computational interest, including those arising from the Clifford hierarchy~\cite{gottesman1999demonstrating}.
These preparations and measurements lie within the basic computational primitives of QCM.
{Our results therefore present a direct operational obstruction for ontological models of QCM that assign overlapping distributions to the state pairs considered:}
such models cannot both satisfy preparation independence and reproduce the relevant Born-rule statistics.
Although stabilizer implementations of particular PBR measurements are already known~\cite{pusey2012reality,karanjai2018contextuality}, the general PBR construction does not impose this measurement restriction.
Our work extends these stabilizer-restricted arguments to computationally relevant families of non-stabilizer states.

Let us summarize our main results.
For the stabilizer fragment, we generalize existing constructions~\cite{pusey2012reality,karanjai2018contextuality} to demonstrate that any pair of distinct nonorthogonal pure stabilizer states yields a PBR proof~(Proposition~\ref{pro:stabilizer-pbr}).
A complementary argument based on contextuality appears in Appendix~\ref{sec:contextuality-pbr}.
Additionally, we construct stabilizer-restricted weakly antidistinguishing measurements for magic-state preparations consisting of tensor products drawn from the single-qubit pairs $R_k\ket+$ and $SR_k\ket+$, where
\[
R_k=\operatorname{diag}(1,e^{i\theta_k}),
\qquad
\theta_k:=\frac{2\pi}{2^k},
\]
and $k\geq3$ labels the non-Clifford levels of the hierarchy~\cite{gottesman1999demonstrating}~(Theorem~\ref{thm:recursive-cyclotomic-pbr}).
The number of qubits used by our construction scales exponentially with the level of the hierarchy according to $D_k=2^{k-2}$.
{The construction uses an algebraic property of these phase-rotated states:} a two-qubit parity measurement maps pairs of such states either to a maximally entangled stabilizer state on the odd-parity branch or to an encoded single-qubit state with phase level reduced by one on the even-parity branch.

It is also possible to construct a stabilizer-restricted PBR proof for every number $n\geq2$ of qubits~(Theorem~\ref{thm:complete-graph-adm}).
For this, we use the preparation pair
\[
\ket{\psi_a}
=
\cos\theta_n\ket0+(-1)^a\sin\theta_n\ket1,
\qquad
a\in\{0,1\},
\qquad
\theta_n=\frac{\pi}{4n},
\]
whose angle depends on the number of qubits.
{The nonadaptive Pauli measurements are specified by the canonical stabilizer generators of the graph state corresponding to $K_n$, i.e. the complete graph on $n$ vertices.}
Each outcome excludes exactly one preparation in the resulting product family.

The above results hold for otherwise arbitrary ontological models reproducing the specified preparation and Pauli-measurement statistics.
We then specialize to the Pauli ontological model, based on the Pauli polytopes introduced in~\cite{zurel2020hidden}.
Within this model, the conceptual issue of preparation independence can be related to a structural question about closure under the operator tensor product.
Our construction of antidistinguishing measurements for the $T$-state pair ($k=3$), together with overlapping ontological representations, demonstrates an obstruction to preparation independence.
In fact, we establish a stronger structural result: the tensor product of any vertex of an $n$-qubit Pauli polytope with any vertex of an $m$-qubit Pauli polytope falls outside the composite $(n+m)$-qubit polytope.
These results obstruct a direct strategy for constructing composite ontological representations by independently sampling and tensoring local vertices.
They therefore preclude a straightforward tensor-product approach to analyzing this simulation scheme, of the kind used in other classical simulation methods~\cite{pashayan2015estimating,howard2017application,bravyi2016improved}, without themselves establishing a lower bound on simulation time.

\subsection{Outline}

The remainder of the paper is organized as follows.
Section~\ref{sec:magic-state-qc} reviews stabilizer theory and quantum computing with magic states.
In {Section~\ref{sec:antidistinguishability} we introduce the operational and ontological frameworks, antidistinguishability, and the preparation-independence argument.}
Sections~\ref{sec:stabilizer-pbr} and~\ref{sec:magic-pbr} develop our antidistinguishing measurements for stabilizer and magic-state preparations, respectively.
In Section~\ref{sec:polytope} we apply these constructions to the Pauli ontological model and establishes the tensor-product obstruction for Pauli-polytope vertices.
We conclude with a discussion of our results in Section~\ref{sec:conclusion}.

\section{Quantum computing with magic states}\label{sec:magic-state-qc}
\subsection{Stabilizer theory}\label{sec:stabilizer-theory}
We begin with some preliminaries on stabilizer theory. A system of $n$ qubits is modeled by a Hilbert space $\mathcal H = \left (\mathbb C^{2}\right )^{\otimes n}$.
An $n$-qubit Pauli operator is any operator of the form
\begin{eqnarray}
i^d\, \sigma_{1}\otimes \cdots \otimes \sigma_{n},\notag
\end{eqnarray}
{where $d\in \ZZ_{4}$ and each $\sigma_i\in\{I,X,Y,Z\}$ is a single-qubit Pauli matrix or the identity}
\[
I = \begin{pmatrix}~1&~0\\~0&~1\end{pmatrix},
\quad
X = \begin{pmatrix}~0&~1\\~1&~0\end{pmatrix},
\quad
Y = \begin{pmatrix}~0&-i\\~i&~0\end{pmatrix},
\quad
Z = \begin{pmatrix}~1&~0\\~0&-1\end{pmatrix}.
\]
For all $n$, Pauli operators either commute or anticommute and are Hermitian if and only if $d = 0,2$.
The $n$-qubit Pauli group is the set of operators
\[
\mathcal G_{n} {:=} \{\pm 1,\pm i\}\times \left \{\sigma_{1}\otimes \cdots {\otimes} \sigma_{n}~:~\sigma_{i}\in \{I,X,Y,Z\}\right \}
\]
with matrix multiplication as the group operation.
A stabilizer group $\mathcal S \subset \mathcal G_{n}$ is an abelian subgroup of the Pauli group not containing $-I$.
Every element of $\mathcal S$ is therefore Hermitian and
squares to the identity.
Such a group has $k$ independent generators,
\[
\mathcal S=\langle s_1,\ldots,s_k\rangle,
\qquad 0\leq k\leq n,
\]
and has order $|\mathcal S|=2^k$.
Note that the generators may include an overall minus sign.
The operator
\[
\Pi_{\mathcal S} := \frac{1}{\lvert \mathcal S \rvert}\sum_{s\in \mathcal S} s
\]
is the projector onto the joint $+1$ eigenspace of each $s\in \mathcal S$.
Measurement of a Pauli observable $A$ obtaining outcome $\sigma\in \{\pm 1\}$ is encoded by the projector $\Pi_{A}^{\sigma} := \Pi_{\mathcal S}$, where $\mathcal S = \langle \sigma A\rangle$.
A stabilizer group $\mathcal S$ is maximal when it has $n$ independent generators. In this case, the projector $\Pi_{\mathcal S} = \ket{\mathcal S}\bra{\mathcal S}$ has rank one and $\ket{\mathcal S}$ is a stabilizer state.

The $n$-qubit Clifford group $\mathcal C_{n}$ is a finite subgroup of the $n$-qubit unitary group $\mathcal U(\mathcal H)$ and, up to phases, is the normalizer of the Pauli group.
Equivalently, the Clifford group is generated by the single-qubit Hadamard $H$ and phase $S$ gates {together with the two-qubit CNOT gate.}
All stabilizer states can be written as
\[
\ket{\mathcal S} = U_{\mathcal S}\ket{0}^{\otimes n},
\]
where $\ket{0}^{\otimes n}$ is a stabilizer state with stabilizer group $\mathcal S = \langle Z_i\rangle_{i=1}^{n}$ and $U_{\mathcal S}$ is a Clifford unitary.
It is known that any stabilizer state can be written in the normal form
\begin{eqnarray}
\ket{\mathcal S} = \frac{1}{\lvert \mathcal A \rvert^{1/2}}\sum_{x\in \mathcal A}{i^{q(x)}}\ket x,%
\label{eq:stab-normal-form}
\end{eqnarray}
where $\mathcal A \subset \ZZ_{2}^{n}$ is an affine subspace and $q:\mathcal A\to \ZZ_{4}$ is a quadratic form; see e.g.,~\cite{van2010classical}.

Any quantum computation initialized in a stabilizer state, then acted on by Clifford gates and measurements of Pauli observables, optionally with classical randomness and feedforward, can be efficiently simulated classically.
This is the well-known Gottesman-Knill theorem~\cite{gottesman1999group22}.

\subsection{Clifford hierarchy and magic state injection}\label{sec:magic-state-injection}
One can promote the stabilizer fragment of quantum theory to a universal quantum computational model with access to non-stabilizer, or `magic' resources.
{A common choice is \(T=\operatorname{diag}(1,e^{i\pi/4})\),}
which amounts geometrically to a $\pi/4$ rotation about the $Z$-axis of the Bloch sphere.
Particularly appealing about the $T$ gate is that it can be implemented in fault-tolerant architectures using state injection schemes~\cite{magic}.

{In this setup, one can implement $T$ deterministically once supplied with a magic state}
\begin{eqnarray}
\ket T
=
T\ket +
=
\frac{1}{\sqrt{2}}\left (\ket 0 + e^{i\pi/4}\ket 1\right ),%
\label{eq:t-state}
\end{eqnarray}
together with adaptive stabilizer operations.
Thus, the primitives for universal quantum computation can be taken to be adaptive Clifford gates and Pauli measurements together with a supply of $\ket T$ magic states.
For drawing samples from a universal family of circuits, the primitives can be further refined to consist only of adaptive Pauli measurements and magic states~\cite{bravyi2016trading}; see also~\cite{yoganathan2019quantum,peres2026normal}.

Other choices of universal gate sets admitting deterministic magic state injection using stabilizer operations are also possible, such as the controlled phase $CS$ and controlled-controlled-$Z$ $CCZ$ gate~\cite{howard2017application}.
{These gates all belong to the third level of the Clifford hierarchy}~\cite{gottesman1999demonstrating}.
More generally, the $k$th-level of the hierarchy is defined recursively by
\[
\mathcal C_{n}^{(k)} := \left \{U\in \mathcal U(\mathcal H)~:~{UAU^{\dagger}\in \mathcal C_{n}^{(k-1)}\text{ for every }A\in\mathcal G_n}\right \},
\]
with the first and second levels being the $n$-qubit Pauli and Clifford groups, respectively.
While the levels of the Clifford hierarchy do not generally form a group, the diagonal elements in each level do form a group.
In particular, at each level $k$ we have the diagonal single-qubit unitary
\begin{eqnarray}
R_{k} := \text{diag}(1,\zeta_{k}),
\quad\text{where}\quad
\zeta_{k} := e^{2\pi i/2^{k}},
\end{eqnarray}
so that $R_{1} = Z$, $R_{2} = S$, and $R_{3} = T$.
Here $\zeta_{k}$ is a root of unity.
The diagonal Clifford hierarchy is naturally described by such roots of unity and phase polynomials~\cite{cui2017diagonal}.

{The corresponding single-qubit resource states can be defined analogously to the $T$ state}
\[
\ket{R_{k}} = R_{k}\ket + = \frac{1}{\sqrt{2}}\left (\ket 0 + \zeta_{k}\ket 1 \right ).
\]
{For $k\geq4$, the standard deterministic injection scheme for $R_k$ additionally requires a lower-level correction $R_{k-1}$ on one measurement branch. This correction can itself be implemented by state injection, with further lower-level corrections as required~\cite{gottesman1999demonstrating}.}
The $n$-fold tensor product of $\ket{R_{k}}$ is given by
\begin{eqnarray}
\ket{R_{k}}^{\otimes n} = 2^{-n/2} \sum_{x\in  \ZZ_{2}^{n}} \zeta_{k}^{\lvert x \rvert}\ket x,%
\label{eq:n-fold-magic-state}
\end{eqnarray}
where $\ZZ_{2}$ is the binary ring and $\lvert x \rvert$ is the Hamming weight of the bit string $x$.
{For $k\geq3$, the coefficients of this normalized state lie in the algebraic field $\QQ(\zeta_k)$. For every $k\geq1$, the coefficients lie in this field after removing the common normalization factor.}
(See \cite{herstein1991topics} as well as Appendix~\ref{sec:field-ext} for further details.)
This fact will prove useful in the developments below.

\section{Antidistinguishability}\label{sec:antidistinguishability}

\subsection{Operational quantum theory}\label{sec:operational-theories}

The primitives of an operational theory are preparation, transformation, and measurement procedures, together with their empirical outcome probabilities.
In quantum theory, preparations are represented by density operators, which are positive operators of unit trace.
Here we consider projective measurements with post-measurement states determined by the L\"uders rule.
In particular, our setup will allow for any quantum state as a valid preparation, but we will focus on a subset of projective measurements $\mathcal M$ with the main example being sequences of Pauli measurements.

Consider a sequence of mutually commuting PVMs $M_j=\{\Pi_j^{k}\}_{k\in K_{j}}$, with $j=1,\ldots,t$, that we denote collectively as $M$.
For an outcome sequence $k =(k_1,\ldots,k_t)$, define the joint projector
\[
\Pi_{M}^{k}:=\prod_{j=1}^t\Pi_j^{k_j}.
\]
The probability of this sequence on a preparation $\rho$ is
\[
p({k}\lvert\rho,M)
=
\Tr(\rho\Pi_{M}^{k}).
\]
Whenever this probability is positive, the corresponding post-measurement state is
\[
\rho_{M,{k}}
=
\frac{\Pi_{M}^{k}\rho\Pi_{M}^{k}}{\Tr(\rho\Pi_{M}^{k})}.
\]
We also allow classical randomization over the permitted procedures. Pure states are the extremal quantum preparations, while rank-one PVMs are the finest projective measurements; higher-rank PVMs can be obtained from them by coarse-graining.

Given a restricted set of measurement procedures $\mathcal M$, two preparations $\rho,\rho'$ are single-shot distinguishable (SSD) if there exists a measurement $M\in\mathcal M$ and a subset $A$ of its outcome set such that
\[
p(A\lvert\rho,M)=1,
\qquad
p(A\lvert\rho',M)=0.
\]
Thus the outcome determines with certainty which preparation was performed.
With access to all PVMs, orthogonal quantum states are always SSD since there exists a PVM that perfectly distinguishes them.

Following~Ref.~\cite{karanjai2018contextuality}, consider now a finite set of pure state preparations $P := \{\rho_s:s\in S\}$.
Performing a measurement $M\in \mathcal M$ and obtaining outcome $k\in K$, we denote the set of post-measurement states by $P_{M,k}$.
The measurement $M$ is said to be a partitioning measurement for $P$ if the set $P_{M,k}$ of {post-measurement states contains} at least one SSD pair for every outcome $k$.

\subsection{Ontological models framework}\label{sec:ontological-models}

Ontological models aim to reproduce empirical statistics using ontic, or \textit{physical}, microstates and classical probability theory.
For simplicity, we take the set of ontic states $\Lambda$ to be finite.%
\footnote{Many of our results extend to general measurable spaces $(\Lambda,\Sigma)$. In that setting, statements about empty intersections of supports are replaced by statements that the preparation measures have no nonzero common submeasure.}
A preparation of $\rho$ is represented by a probability distribution $\mu_\rho$ on  $\Lambda$.
For two preparation distributions $\mu_0$ and $\mu_1$,
define their common subdistribution by
\[
\nu(\lambda):=\min\{\mu_0(\lambda),\mu_1(\lambda)\},
\]
and their overlap by
\[
\omega(\mu_0,\mu_1):=\sum_{\lambda\in\Lambda}\nu(\lambda).
\]
{The support of a distribution $\mu$ is defined as}
\[
\Supp(\mu):=\{\lambda\in\Lambda:\mu(\lambda)>0\}.
\]
{Drawing on standard terminology, we say that an ontological model is $\psi$-ontic if}
$\omega(\mu_\rho,\mu_{\rho'})=0$ for every pair of distinct pure-state preparations $\rho$ and $\rho'$.
{Otherwise, it is $\psi$-epistemic~\cite{harrigan2010einstein}.}

A projective measurement $M$ with outcome set $K$ is represented by a transition kernel
\[
\Gamma(k,\lambda'\lvert M,\lambda)\geq 0,
\qquad
\sum_{k\in K}\sum_{\lambda'\in\Lambda}
\Gamma(k,\lambda'\lvert M,\lambda)=1
\]
for every $\lambda\in\Lambda$.
This kernel specifies the joint probability of obtaining outcome $k$ and transitioning to $\lambda'$, conditional on the initial ontic state $\lambda$.
The measurement response functions are obtained by marginalizing over the final ontic state:
\[
\xi(k\lvert M,\lambda)
:=\sum_{\lambda'\in\Lambda}
\Gamma(k,\lambda'\lvert M,\lambda)
\]
and the model reproduces the operational probabilities when
\[
p(k\lvert\rho,M)
=\sum_{\lambda\in\Lambda}
\mu_\rho(\lambda)\,\xi(k\lvert M,\lambda).
\]
Whenever this probability is positive, the conditional post-measurement distribution is
\[
\mu_{\rho;M,k}(\lambda')
=
\frac{\sum_{\lambda\in\Lambda}
\Gamma(k,\lambda'\lvert M,\lambda)\mu_\rho(\lambda)}
     {p(k\lvert\rho,M)}.
\]
This distribution must reproduce the subsequent measurement statistics of the corresponding L\"uders post-measurement state.

\subsection{Antidistinguishing measurements}\label{sec:ad-measurement}

For preparations that are SSD, the measurement procedure $M$ that achieves this separation is called a distinguishing measurement. We will now consider antidistinguishing measurements (ADMs), which are measurements whose outcomes certify that some preparations were definitely not used in that run of the experiment, but typically do not confirm which preparation was actually used.

Consider a finite preparation family $\{\rho_s:s\in S\}$ with measurement set $\mathcal M$.
Following~\cite{srikumar2026contextuality}, we say that a projective measurement $M=\{\Pi_M^k:k\in K\}\in\mathcal M$ is a weakly \textit{antidistinguishing} measurement (WADM) for this family if there exists a map $f:K\to S$ such that
\[
p(k\lvert\rho_{f(k)},M)=0
\qquad\text{for every }k\in K.
\]
Equivalently, every measurement outcome is excluded by at least one member of the preparation family.
We say $M$ is a strongly antidistinguishing measurement (SADM) if it is a WADM and, for every $s\in S$, there exists an outcome $k_s\in K$ such that
\[
p(k_s\lvert\rho_s,M)=0,
\qquad
p(k_s\lvert\rho_{s'},M)>0
\quad\text{for all }s'\ne s.
\]
Since every SADM is also a WADM, we establish our results for the weak case unless stated otherwise.
{We will often use ADM simply as an abbreviation broadly referring to antidistinguishing measurement without invoking the specific qualifiers above.}

\begin{thm}\label{thm:restricted-pbr}
If a finite preparation family $\{\rho_s:s\in S\}$ admits a WADM,
then every ontological model reproducing its zero probabilities satisfies
\[
\bigcap_{s\in S}\Supp(\mu_{\rho_s})=\emptyset.
\]
\end{thm}

\begin{proof}
Suppose that $\lambda$ belongs to the support of every $\mu_{\rho_s}$.
For each outcome $k$, the WADM condition gives
\[
p(k\lvert\rho_{f(k)},M)
=
\sum_{\lambda'\in\Lambda}\mu_{\rho_{f(k)}}(\lambda')\,\xi(k\lvert M,\lambda')
=
0.
\]
Every summand is nonnegative. Since $\mu_{\rho_{f(k)}}(\lambda)>0$, it follows that
$\xi(k\lvert M,\lambda)=0$.
This holds for every $k$, contradicting $\sum_{k\in K}\xi(k\lvert M,\lambda)=1$.
\end{proof}

Let $\rho_0,\rho_1\in\mathcal P$ be two preparations.
For $q\geq1$, define their complete product family by
\[
\rho_x:=\rho_{x_1}\otimes\cdots\otimes\rho_{x_q},
\qquad x\in\ZZ_2^q.
\]
The ontological model satisfies \textit{preparation independence} for this family if its composite ontic space is $\Lambda^q$ and the preparation is represented by a product distribution
\[
\mu_{\rho_x}(\lambda_1,\ldots,\lambda_q)
=\prod_{j=1}^q\mu_{\rho_{x_j}}(\lambda_j).
\]

\begin{cor}[PBR argument~\cite{pusey2012reality}]
\label{cor:restricted-pbr}
Suppose the complete $q$-copy product family generated by $\rho_0,\rho_1$ admits a WADM.
If an ontological model reproduces its zero probabilities and satisfies preparation independence, then
\[
\Supp(\mu_{\rho_0})\cap\Supp(\mu_{\rho_1})=\emptyset,
\]
or equivalently, $\omega(\mu_{\rho_0},\mu_{\rho_1})=0$.
\end{cor}

\begin{proof}
Suppose that
$\lambda\in\Supp(\mu_{\rho_0})\cap\Supp(\mu_{\rho_1})$.
Preparation independence implies that, for every $x\in\ZZ_2^q$,
\[
\mu_{\rho_x}(\lambda,\ldots,\lambda)
=\prod_{j=1}^q\mu_{\rho_{x_j}}(\lambda)>0.
\]
Thus $(\lambda,\ldots,\lambda)$ belongs to the common support of the entire product family, contradicting Theorem~\ref{thm:restricted-pbr}.
\end{proof}

We remark that the PBR argument holds if and only if you regard preparation independence as fundamental.
One could instead relinquish this principle by introducing additional `hidden' correlations in the ontological representation of preparations, similar in spirit to how Bell's theorem forces nonlocality in ontological models.

\begin{pro}\label{pro:partitioning-wadm}
{If a preparation family $\{\rho_s:s\in S\}$ admits a partitioning measurement whose conditional distinguishing measurements commute with its projectors, and the resulting adaptive measurement is allowed, then it admits a WADM.}
\end{pro}

\begin{proof}
Let $M$ be a partitioning measurement for
$\{\rho_s:s\in S\}$.
For each outcome $k\in K$, choose two preparations
$\rho_s,\rho_{s'}$ whose conditional post-measurement states
$\rho_{s;M,k},\rho_{s';M,k}$ are SSD.
Here the indices $s,s'$ may depend on $k$.
Conditional on $k$, perform an allowed measurement $N_k$
that distinguishes these states.
Writing $R_k$ for its outcome set, there exists a subset
$A_k\subset R_k$, with complement
$A_k^\perp:=R_k\backslash A_k$, such that
\[
p(A_k^\perp\lvert\rho_{s;M,k},N_k)=0,
\qquad
p(A_k\lvert\rho_{s';M,k},N_k)=0.
\]
Let $\widetilde M$ denote the resulting adaptive measurement,
whose outcomes are pairs $(k,r)$ with $r\in R_k$.
The joint probabilities satisfy
\[
p(k,r\lvert\rho_s,\widetilde M)
=
p(k\lvert\rho_s,M)\,
p(r\lvert\rho_{s;M,k},N_k),
\]
whenever the first factor is positive; otherwise the joint
probability is zero.
Consequently,
\[
p(k,r\lvert\rho_{s'},\widetilde M)=0
\quad\text{if }r\in A_k,
\]
while
\[
p(k,r\lvert\rho_s,\widetilde M)=0
\quad\text{if }r\in A_k^\perp.
\]
{Writing $P_k$ for the projectors of $M$ and $Q_{r|k}$ for those of $N_k$, the joint effects are $P_kQ_{r|k}P_k=P_kQ_{r|k}$. By the commutation assumption, these are mutually orthogonal projectors summing to the identity. Thus $\widetilde M$ is a PVM, not necessarily of rank one.}
Every outcome of $\widetilde M$ therefore excludes at least
one original preparation, so $\widetilde M$ is a WADM.
\end{proof}

\section{Antidistinguishing measurements for stabilizer states}\label{sec:stabilizer-pbr}

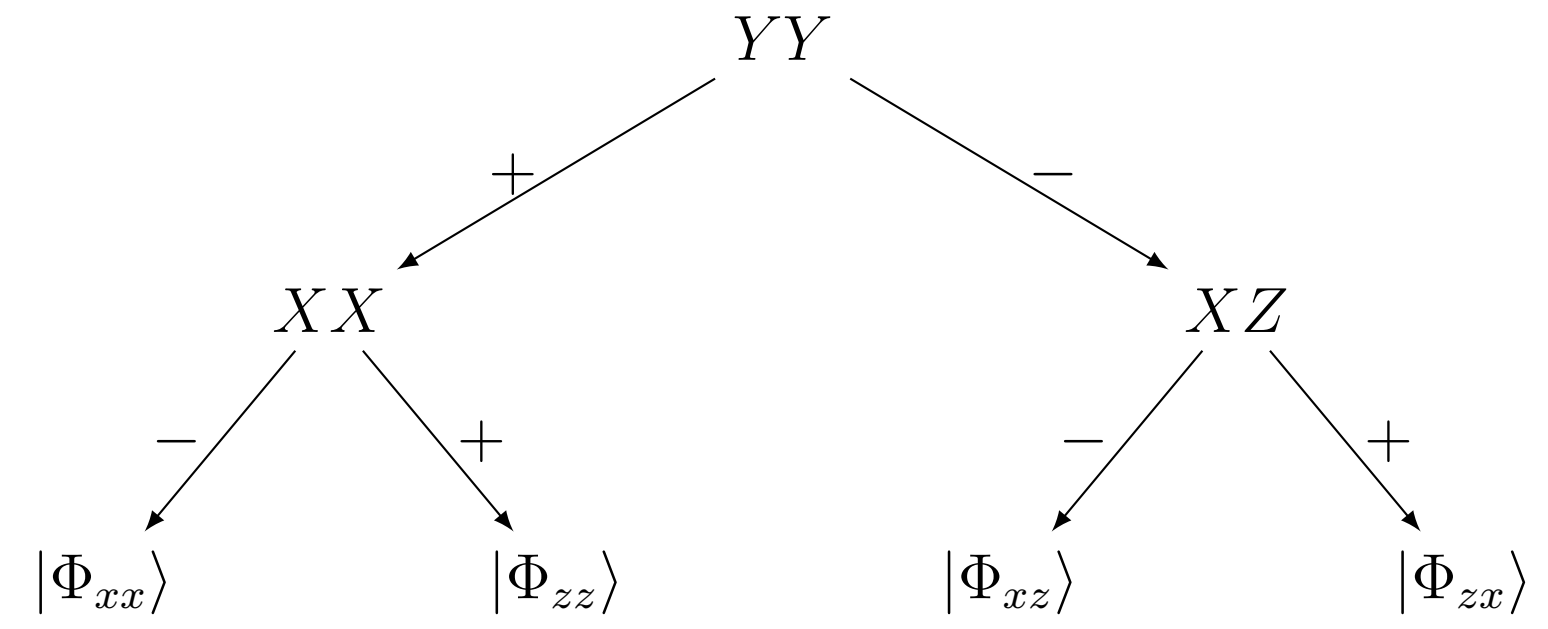
\begin{figure}[h]
\centering
\begin{tikzpicture}[
  grow=down,
  level distance=18mm,
  level 1/.style={sibling distance=60mm},
  level 2/.style={sibling distance=30mm},
  edge from parent/.style={draw,-latex},
  every node/.style={font=\large}
]

\node {$YY$}
  child {
    node {$XX$}
      child { node {$\ket{\Phi_{xx}}$}
              edge from parent node[left] {${-}$} }
      child { node {$\ket{\Phi_{zz}}$}
              edge from parent node[right] {${+}$} }
    edge from parent node[left] {$+$}
  }
  child {
    node {$XZ$}
      child { node {$\ket{\Phi_{xz}}$}
              edge from parent node[left] {${-}$} }
      child { node {$\ket{\Phi_{zx}}$}
              edge from parent node[right] {${+}$} }
    edge from parent node[right] {$-$}
  };

\end{tikzpicture}
\caption{The PVM in Eqns.~(\ref{eq:pbr-pvm-xx})-(\ref{eq:pbr-pvm-zz}) is implemented by an adaptive measurement tree: measure $YY$ first, {measure $XX$ on the $+1$ branch, and measure $XZ$ on the $-1$ branch.} Since each terminal outcome is orthogonal to its correspondingly labelled product preparation it is a SADM for these preparations.}
\label{fig:pbc-tree}
\end{figure}

Simple examples of ADMs restricted to Pauli measurements and stabilizer state preparations go back to the original work of PBR~\cite{pusey2012reality}; see also~\cite{karanjai2018contextuality}.
{Consider the nonorthogonal single-qubit stabilizer states}
$$\ket{\psi_{x}} := \frac{1}{\sqrt{2}}(\ket 0 + \ket 1)\quad \text{and}\quad \ket{\psi_{z}} := \ket 0,$$
as well as the two-qubit state vectors
\begin{subequations}
\begin{eqnarray}
\ket{\Phi_{xx}}
&=&
\frac{\ket{00} - \ket{11}}{\sqrt{2}},\label{eq:pbr-pvm-xx}\\
\ket{\Phi_{xz}}
&=&
{\frac{\ket{00} + \ket{01} -\ket{10} + \ket{11}}{2}},\label{eq:pbr-pvm-xz}\\
\ket{\Phi_{zx}}
&=&
{\frac{\ket{00} - \ket{01} +\ket{10} + \ket{11}}{2}},\label{eq:pbr-pvm-zx}\\
{\ket{\Phi_{zz}}}
&=&
\frac{\ket{01} + \ket{10}}{\sqrt{2}}.\label{eq:pbr-pvm-zz}
\end{eqnarray}
\end{subequations}
{These rays form a PVM that can be implemented by the adaptive measurement sequence pictured} in Fig.~\ref{fig:pbc-tree}.
By direct calculation we have
\begin{eqnarray}
\lvert \bra{\Phi_{ab}}\left (\ket{\psi_{a}}\otimes \ket{\psi_{b}}\right )\rvert^{2} = 0,
\quad \text{and} \quad
\lvert \bra{\Phi_{ab}}\left (\ket{\psi_{c}}\otimes \ket{\psi_{d}}\right )\rvert^{2} > 0,
\quad ab \neq cd.%
\notag
\end{eqnarray}
Consequently, this is a SADM, which by Corollary~\ref{cor:restricted-pbr} implies that a PBR proof exists in this case.

{This example can be generalized to arbitrary nonorthogonal stabilizer states} $\ket{\mathcal S_{0}}$ and $\ket{\mathcal S_{1}}$.
Our argument relies on the following standard result.

\begin{lem}\label{lem:stabilizer-pair-normal-form}
Let $\ket{\mathcal S_0}$ and $\ket{\mathcal S_1}$ be distinct, non-orthogonal pure $n$-qubit stabilizer states.  There are a Clifford unitary $C$, an integer $r\geq1$, and a relabelling of tensor factors such that
\[
C\ket{\mathcal S_0}=\ket{0}^{\otimes r}\otimes\ket{0}^{\otimes(n-r)},\qquad
C\ket{\mathcal S_1}=\ket{+}^{\otimes r}\otimes\ket{0}^{\otimes(n-r)}.
\]
\end{lem}

\begin{proof}
First send $\ket{\mathcal S_0}$ to $\ket{0}^{\otimes n}$ by a Clifford $C$.
From the stabilizer state normal form Eq.~(\ref{eq:stab-normal-form}), we see that for the transformed ${C\ket{\mathcal S_1}}$ to be non-orthogonal that the corresponding affine subspace is, in fact, a linear subspace containing the origin.
We can perform row reduction on a basis of this linear subspace using CNOT and SWAP gates, which fix $\ket{0}^{\otimes n}$.
Diagonal Clifford gates remove the remaining linear and quadratic phases, again while fixing $\ket{0}^{\otimes n}$ up to phase.
The second state is then $\ket{+}^{\otimes r}\ket{0}^{\otimes(n-r)}$.
Requiring that the states be distinct gives $r\geq1$.
\end{proof}

{The generalization then proceeds by choosing an odd number $m$ of active qubits in each block: take $m=r$ for odd $r$ and $m=r-1$ for even $r$. The operators $\overline X=X^{\otimes m}$ and $\overline Z=Z^{\otimes m}$, acting trivially on the remaining qubits, anticommute and define logical Pauli operators, with $\overline Y=i\overline X\overline Z$. In the corresponding logical-qubit subsystem, $\ket0^{\otimes r}$ has logical state $\ket0$ and $\ket+^{\otimes r}$ has logical state $\ket+$; their states on the remaining subsystem need not coincide. The common stabilizer tail $\ket0^{\otimes(n-r)}$ does not enter the measurement.
We then perform the logical two-qubit adaptive Pauli PVM of Figure~\ref{fig:pbc-tree} on the two preparation blocks. Its observables commute along each branch, so its terminal effects are mutually orthogonal projectors, which need not have rank one.} 

\begin{pro}\label{pro:stabilizer-pbr}
An ontological model reproducing the predictions of stabilizer theory and {satisfying preparation independence} must represent distinct stabilizer states $\ket{\mathcal S_{0}}$ and $\ket{\mathcal S_{1}}$ by probability distributions $\mu_{0}$ and $\mu_{1}$ with disjoint support, i.e. $\text{Supp}(\mu_{0})\cap \text{Supp}(\mu_{1}) = \emptyset$.
\end{pro}

{The result holds since every terminal outcome of the logical Pauli PVM excludes one of the four product preparations. Orthogonal pairs are already distinguishable by a Pauli measurement.}
For a contextuality-based proof of this result, see Appendix~\ref{sec:contextuality-pbr};
similar connections between preparation independence and the contextuality of the stabilizer fragment were also explored in~\cite{lillystone2019contextual}.

{One straightforward consequence of Proposition~\ref{pro:stabilizer-pbr}} is that an ontological model reproducing the predictions of qubit stabilizer theory,
but that does not represent non-orthogonal stabilizer states by \textit{disjoint} probability distributions,
must then violate preparation independence.
An example of such a model is the closed non-contextual (CNC) phase space introduced in \cite{raussendorf2020phase} (see also \cite{ipek2025phase}),
{which provides an operator-based ontological model} for qubit stabilizer quantum mechanics.
These operators, however, are not closed under tensor product~\cite{okay2021extremal} so that preparation independence fails in this model.
We prove a generalization of this result in Section~\ref{sec:lambda-application} for an operator-based ontological model of universal QCM.

\section{Antidistinguishing measurements for magic states}
\label{sec:magic-pbr}

The input resource state for QCM is often taken to be a tensor
product of magic states.
In this section we construct WADMs for non-stabilizer input states.
For convenience, we will often drop normalization factors, as
they do not affect {the zero-probability condition.}

\subsection{Partitioning measurement for third-level states}%
\label{sec:partitioning-t-states}

Here we establish WADMs for tensor products of $T$ magic states
by constructing a partitioning measurement.
Let $\ket{T_a}:=Z^a\ket T$, where $Z\ket T$ is
orthogonal to $\ket T$.
Define now
\[
\ket{\psi_{ab}}
:=
\ket{T_a}\otimes\ket{T_b},
\qquad a,b\in\ZZ_2,
\]
or more explicitly
\[
\ket{\psi_{ab}}
{\propto}
\ket{00}+(-1)^{a+b}i\ket{11}
+{\zeta_3}\left(
(-1)^b\ket{01}+(-1)^a\ket{10}
\right).
\]
Consider now the parity measurement of the two-qubit Pauli $ZZ$
yielding the post-measurement states
{represented, up to phase and normalization, by}
\begin{eqnarray}
\ket{u_E(a+b)}
&:=&
\ket{00}+(-1)^{a+b}i\ket{11},
\label{eq:k_3-projected-1}\\
\ket{u_O(a+b)}
&:=&
\ket{01}+(-1)^{a+b}\ket{10},
\label{eq:k_3-projected-2}
\end{eqnarray}
where $E$ and $O$ correspond to the even and odd parity
subspaces, respectively.
These are, in fact, stabilizer states satisfying
\[
XY\ket{u_E(a+b)}
=(-1)^{a+b}\ket{u_E(a+b)},
\qquad
XX\ket{u_O(a+b)}
=(-1)^{a+b}\ket{u_O(a+b)}.
\]
Thus, conditional on the outcome $ZZ=+1$, measure $XY$;
conditional on $ZZ=-1$, measure $XX$.
Each conditional measurement distinguishes the two possible
post-measurement states.
The Pauli $ZZ$ is therefore a partitioning measurement for this
set of states and the conditional measurements $XY$ and $XX$
extend this to a WADM.
This is precisely the adaptive measurement sequence in
Fig.~\ref{fig:pbc-tree}, up to single-qubit Clifford unitaries.

\begin{pro}\label{pro:t-block-zero-cover}
For every $n\geq2$, the preparation family
\[
\rho_x:=\ket{T_x}\bra{T_x},
\qquad
\ket{T_x}:=\bigotimes_{\ell=1}^{n}\ket{T_{x_\ell}},
\qquad x\in\ZZ_2^n,
\]
admits a rank-one WADM implemented by adaptive Pauli
measurements.
\end{pro}

\begin{proof}
Partition the qubits into disjoint pairs.
On each pair $(i,j)$, measure $Z_iZ_j$, followed by
$X_iY_j$ on the $+1$ branch and $X_iX_j$ on the $-1$ branch.
If $n$ is odd, measure $Z$ on the remaining qubit.
The preceding two-qubit calculation shows that the second
outcome on each pair is deterministically $(-1)^{x_i+x_j}$.
Thus, every complete outcome string excludes preparations
with the opposite parity on any pair, establishing the
WADM property.
We see that along each branch, the protocol measures $n$
independent commuting Pauli observables, whose joint
eigenspaces have rank one.
Distinct branches give orthogonal projectors,
{since their first differing outcomes belong to orthogonal
eigenspaces of the same observable}.
Thus these projectors form a rank-one PVM.
\end{proof}

Although $\ket T$ and $Z\ket T$ are orthogonal, they are not
SSD by a single Pauli measurement, and so their ontological
representations could, in principle, overlap in this
restricted setting.
The WADM constructed above, together with
Corollary~\ref{cor:restricted-pbr}, however, rules this out
under preparation independence.

For later developments, the same protocol also gives a
conclusion assuming independence only between two blocks.
{We write the block states explicitly as
$\ket T^{\otimes n}$ and $(Z\ket T)^{\otimes n}$.}

\begin{cor}[Block preparation independence]
\label{cor:magic-pi}
For $n,n'\geq1$, let $\mu_n$, $\mu_{n'}$, and
$\mu_{n'}^\perp$ represent
{$\ket T^{\otimes n}$, $\ket T^{\otimes n'}$, and
$(Z\ket T)^{\otimes n'}$}, respectively.
Suppose an ontological model reproduces the adaptive Pauli
protocol above and represents the composite preparations
{
\[
\ket T^{\otimes n}\otimes\ket T^{\otimes n'},
\qquad
\ket T^{\otimes n}\otimes(Z\ket T)^{\otimes n'}
\]
}
by $\mu_n\otimes\mu_{n'}$ and
$\mu_n\otimes\mu_{n'}^\perp$, respectively.
Then
\[
\Supp(\mu_{n'})\cap\Supp(\mu_{n'}^\perp)=\emptyset,
\]
or equivalently,
$\omega(\mu_{n'},\mu_{n'}^\perp)=0$.
Preparation independence is not assumed within either block.
\end{cor}

\begin{proof}
Choose $i$ in the first block and $j$ in the second.
For the two preparations, the selected qubits are in
$\ket{\psi_{00}}$ and $\ket{\psi_{01}}$,
{respectively. The two preparations can also differ on the
unmeasured qubits, but these do not affect the selected-qubit probabilities.}
The two-qubit {adaptive protocol} described above therefore
gives opposite deterministic second outcomes.
Every terminal outcome excludes one of the two preparations,
so the protocol is a WADM for this pair.
{Theorem~\ref{thm:restricted-pbr} gives
\[
\Supp(\mu_n\otimes\mu_{n'})
\cap
\Supp(\mu_n\otimes\mu_{n'}^\perp)
=\emptyset.
\]
The intersection equals
\[
\Supp(\mu_n)\times
\bigl(\Supp(\mu_{n'})\cap\Supp(\mu_{n'}^\perp)\bigr).
\]
Since $\mu_n$ is normalized, its support is nonempty,
and the result follows.}
\end{proof}

\subsection{A hierarchy of antidistinguishing measurements}
\label{sec:cyclotomic-pbr}

We now generalize the previous construction, based on the $ZZ$
parity measurements, to construct {WADMs} for additional
families of preparations arising from higher levels of the
Clifford hierarchy.
Define the single-qubit states
\begin{equation}
\ket{\psi_b^{(k)}}
:=
S^bR_k\ket+
=
{
\frac{1}{\sqrt2}
\left(\ket0+i^b\zeta_k\ket1\right),
\qquad b\in\ZZ_2,
}
\label{eq:cyclotomic-atoms}
\end{equation}
where $R_k$ and $\zeta_k$ are as in
Section~\ref{sec:magic-state-qc}.
{Recall that $\zeta_k=e^{2\pi i/2^k}$.}
{The two states are nonorthogonal for every $k\geq2$.}
To each $n$-bit string $z\in\ZZ_2^n$, we associate the
preparation
\[
\ket{\psi_z^{(k)}}
:=
\bigotimes_{i=1}^n\ket{\psi_{z_i}^{(k)}}
=
{2^{-n/2}}
\sum_{x\in\ZZ_2^n}
i^{z\cdot x}\zeta_k^{|x|}\ket x,
\]
where {$z\cdot x:=\sum_i z_ix_i$ is evaluated over the
integers so that the exponent of $i$ is evaluated modulo four, not modulo two.}
Let $\ket\phi$ be a stabilizer state.
Using the normal form in Eq.~(\ref{eq:stab-normal-form}),
we have the overlap
{
\begin{equation}
\langle\phi\lvert\psi_z^{(k)}\rangle
=
{\frac{P_{\phi,z}(\zeta_k)}{\sqrt{2^n|\mathcal A|}}},
\quad\text{where}\quad
P_{\phi,z}(t)
:=
{\sum_{x\in\mathcal A}i^{z\cdot x-q(x)}t^{|x|}}.
\label{eq:cyclotomic-overlap}
\end{equation}
}
We will be interested in when these overlaps are precisely zero,
since this corresponds to the outcome of a stabilizer PVM
having zero probability.

To this end, it will be useful to interpret this overlap
{through the polynomial} $P_{\phi,z}(t)$ in a complex
variable $t$ with coefficients in
\[
\QQ(i):=\{a+ib:a,b\in\QQ\},
\]
which is {a field}~\cite{herstein1991topics};
see also Appendix~\ref{sec:field-ext}.
This polynomial has degree at most $n$.
At the same time, the minimal polynomial of $\zeta_k$ over
$\QQ(i)$ is given by
\begin{equation}
m_k(t)=t^{D_k}-i,
\label{eq:minimal-cyclotomic-poly}
\end{equation}
which has degree $D_k:=2^{k-2}$.
These algebraic developments are important for making the
following observation.

\begin{obs}\label{pro:field-degree-threshold}
If $n<D_k$, then a stabilizer overlap
in~\eqref{eq:cyclotomic-overlap} vanishes if and only if
$P_{\phi,z}(t)$ is the zero polynomial.
Equivalently, every Hamming-weight coefficient vanishes
separately.
In particular, the first qubit number at which a zero can
use the new algebraic relation $\zeta_k^{D_k}=i$ is $n=D_k$.
\end{obs}

{The threshold establishes the minimal number of qubits
at which a nonzero overlap polynomial can vanish at $\zeta_k$.
Below the threshold, zero overlaps can still occur through
cancellation within each Hamming-weight coefficient.}
Indeed, at the threshold $n=D_k$, we have the cat state
{
\[
\ket{\mathcal S_1^{(k)}}
=
\frac{1}{\sqrt2}
\left(\ket0^{\otimes D_k}-i\ket1^{\otimes D_k}\right),
\]
}
with stabilizer group
{
\[
\mathcal S_1^{(k)}
=
\left\langle
-Y_1X_2\cdots X_{D_k},
\,Z_iZ_{i+1}:i=1,\ldots,D_k-1
\right\rangle.
\]
}
This state gives an \textit{algebraic} zero
{
\[
\langle\mathcal S_1^{(k)}\lvert
(\psi_0^{(k)})^{\otimes D_k}\rangle
=
2^{-(D_k+1)/2}\left(1+i\zeta_k^{D_k}\right)=0.
\]
The threshold alone does not establish the minimum number
of qubits needed for a WADM. At the threshold, however,
we have the following result for every $k\geq3$.
}

\begin{thm}\label{thm:recursive-cyclotomic-pbr}
{For every $k\geq3$, the complete product family generated
by the nonorthogonal single-qubit states
in~\eqref{eq:cyclotomic-atoms} admits a rank-one stabilizer
WADM on
\[
D_k=2^{k-2}
\]
independently prepared copies.}
\end{thm}

The key observation is that a parity measurement acting on
$\ket{\psi_a^{(k)}}\otimes\ket{\psi_b^{(k)}}$ yields the
even and odd parity post-measurement states
{represented, up to phase and normalization, by}
{
\begin{eqnarray}
\ket{v_E^{(k)}(a,b)}
&:=&
\ket{00}+i^{a+b}\zeta_k^2\ket{11},
\label{eq:hierarchy-projected-E}\\
\ket{v_O^{(k)}(a,b)}
&:=&
\ket{01}+i^{a-b}\ket{10}.
\label{eq:hierarchy-projected-O}
\end{eqnarray}
}
The odd case yields a stabilizer state.
In the even case, since
\[
\zeta_{k-1}=\zeta_k^2,
\]
we recognize the post-measurement state as
{an encoded single qubit with phase level $k-1$}.
Here the encoding is $\ket0\mapsto\ket{00}$ and
$\ket1\mapsto\ket{11}$.
{Note that the exponent $a+b$ is an integer sum, so it may equal two.
For the recursive construction, we use the inputs $00$
and $01$, which give the encoded states
$\ket{\psi_0^{(k-1)}}$ and $\ket{\psi_1^{(k-1)}}$,
respectively.}
Our proof will use this recursive property to construct a WADM.

{To see how this works, let us work out the cases $k=3,4$,}
the first two nontrivial instances.
For $k=3$ we {use} $D_3=2$ qubits.
Let $z=ab$ label the product preparation
$\ket{\psi_{ab}^{(3)}}$.
Performing first the parity measurement $ZZ$ we obtain
\begin{eqnarray}
\ket{v_E^{(3)}(a,b)}
&:=&
\ket{00}+i^{a+b+1}\ket{11},
\label{eq:k_3-projected-E}\\
\ket{v_O^{(3)}(a,b)}
&:=&
\ket{01}+i^{a-b}\ket{10}.
\label{eq:k_3-projected-O}
\end{eqnarray}
Notice that both post-measurement states are stabilizer states
in {this} case.
Measure next the Pauli $XY$, regardless of outcome.
Equivalently, we perform the stabilizer PVM
{with rays}
{
\begin{align}
\ket{\Omega_{00}}&=\ket{00}-i\ket{11},
&
\ket{\Omega_{11}}&=\ket{00}+i\ket{11},
\notag\\
\ket{\Omega_{01}}&=\ket{01}+i\ket{10},
&
\ket{\Omega_{10}}&=\ket{01}-i\ket{10}.
\label{eq:k3-adm-rays}
\end{align}
}
It is straightforward to check that each vector is an
eigenstate of both $ZZ$ and $XY$.
The labeling is chosen such that the preparation
$\ket{\psi_{ab}^{(3)}}$ is excluded by the PVM element
{projecting onto $\ket{\Omega_{ab}}$}.

The higher-level case of $k=4$ will build {on the}
lower-level measurement scheme.
Now we {use} $D_4=4$ qubits.
{Recall that $\zeta_4=e^{i\pi/8}$ and
$\zeta_4^2=\zeta_3=e^{i\pi/4}$.}
The bit string $z=z_1z_2z_3z_4$ corresponds to the product
preparation
\[
{\ket{\psi_z^{(4)}}}
=
\ket{\psi_{z_1}^{(4)}}\otimes\cdots\otimes
\ket{\psi_{z_4}^{(4)}}.
\]
Pair qubits $(1,2)$ and $(3,4)$ and first measure
$Z_1Z_2$ and $Z_3Z_4$.
Introduce the even and odd code bases
\[
\ket{0_E}=\ket{00},\quad \ket{1_E}=\ket{11},
\qquad
\ket{0_O}=\ket{01},\quad \ket{1_O}=\ket{10},
\]
together with the logical eigenstates, for
$\sigma\in\{+1,-1\}$,
\[
\ket{x_\ell^\sigma}
=\ket{0_\ell}+\sigma\ket{1_\ell},
\qquad
\ket{y_\ell^\sigma}
=\ket{0_\ell}+\sigma i\ket{1_\ell},
\qquad \ell\in\{E,O\}.
\]
After the parity measurement on either pair, the projected
two-qubit states are
\begin{align}
\ket{v_E(b,c)}
&=\ket{0_E}+i^{b+c}{\zeta_3}\ket{1_E},
&
\ket{v_O(b,c)}
&=\ket{0_O}+i^{b-c}\ket{1_O}.
\label{eq:sqrtT-four-pair-states}
\end{align}
{For inputs $bc=00,01$, the even pair forms an encoded
version of $\ket{\psi_0^{(3)}}$ or
$\ket{\psi_1^{(3)}}$, respectively, one level lower in
the hierarchy.}
On an odd pair we have that the string $bc=01$ corresponds
to the stabilizer state $\ket{y_O^-}$;
similarly, $10$ to $\ket{y_O^+}$, and both $00$ and $11$
correspond to $\ket{x_O^+}$.

The initial two parity measurements produce the syndrome
$(\ell,\ell')$, where $\ell,\ell'{\in\{E,O\}}$.
Whenever we obtain an odd parity syndrome outcome there
is a qubit pair in a stabilizer state
{on which a logical $Y$ measurement excludes input $01$
for one outcome and input $10$ for the other}.
The $(E,E)$ syndrome is the recursive branch since,
as mentioned above,
{the selected inputs $00,01$ on each even pair give
encoded single-qubit third-level states}.
Our stabilizer PVM will use the four stabilizer rays
\begin{align}
\ket{\Xi_{00}}
&=\frac{\ket{0000}-i\ket{1111}}{\sqrt2},
&
\ket{\Xi_{11}}
&=\frac{\ket{0000}+i\ket{1111}}{\sqrt2},
\notag\\
\ket{\Xi_{01}}
&=\frac{\ket{0011}+i\ket{1100}}{\sqrt2},
&
\ket{\Xi_{10}}
&=\frac{\ket{0011}-i\ket{1100}}{\sqrt2}.
\label{eq:sqrtT-four-even-rays}
\end{align}
These are exactly the encoded versions of the
{measurement vectors $\ket{\Omega_{ab}}$ in
Eq.~\eqref{eq:k3-adm-rays}}.
{All four syndrome subspaces are refined to rank-one
elements by the following independent, conditional Pauli
measurements:}
\begin{equation}
\begin{array}{c|c}
\text{syndrome}&\text{conditional Pauli generators}\\ \hline
EE&Z_1Z_3,\quad X_1X_2Y_3X_4\\
EO&X_1X_2,\quad Y_3X_4\\
OE&Y_1X_2,\quad X_3X_4\\
OO&Y_1X_2,\quad X_3X_4
\end{array}
\label{eq:sqrtT-four-pauli-tree}
\end{equation}
{The last three rows} produce the remaining orthogonal vectors
\begin{align}
\mathcal B_{EO}
&=\{\ket{x_E^t}_{12}\ket{y_O^s}_{34}:s,t=\pm1\},
\notag\\
\mathcal B_{OE}
&=\{\ket{y_O^s}_{12}\ket{x_E^t}_{34}:s,t=\pm1\},
\notag\\
\mathcal B_{OO}
&=\{\ket{y_O^s}_{12}\ket{x_O^t}_{34}:s,t=\pm1\}.
\label{eq:sqrtT-four-other-rays}
\end{align}
Altogether, the four sets in~\eqref{eq:sqrtT-four-even-rays}
and~\eqref{eq:sqrtT-four-other-rays} give the
sixteen-ray stabilizer PVM.
In Table~\ref{tab:sqrtT-four-zero-assignment} we identify
a bijection between preparations and the corresponding
impossible outcome.
In the non-even rows, the first ket refers to qubits $(1,2)$
and the second to qubits $(3,4)$.
Every one of the sixteen preparation words occurs exactly once.

\begin{table}[H]
\centering
\small
{
\begin{tabular}{@{}cll@{}}
\toprule
Syndrome & PVM ray & Preparation $z$\\
\midrule
$EE$ & $\Xi_{00}$ & $0000$\\
$EE$ & $\Xi_{01}$ & $0001$\\
$EE$ & $\Xi_{10}$ & $0100$\\
$EE$ & $\Xi_{11}$ & $0011$\\
\midrule
$EO$ & $x_E^+\otimes y_O^+$ & $1001$\\
$EO$ & $x_E^-\otimes y_O^+$ & $1101$\\
$EO$ & $x_E^+\otimes y_O^-$ & $0010$\\
$EO$ & $x_E^-\otimes y_O^-$ & $1110$\\
\midrule
$OE$ & $y_O^+\otimes x_E^+$ & $0101$\\
$OE$ & $y_O^+\otimes x_E^-$ & $0110$\\
$OE$ & $y_O^-\otimes x_E^+$ & $1000$\\
$OE$ & $y_O^-\otimes x_E^-$ & $1010$\\
\midrule
$OO$ & $y_O^+\otimes x_O^+$ & $0111$\\
$OO$ & $y_O^+\otimes x_O^-$ & $1100$\\
$OO$ & $y_O^-\otimes x_O^+$ & $1011$\\
$OO$ & $y_O^-\otimes x_O^-$ & $1111$\\
\bottomrule
\end{tabular}
}
\caption{Each product preparation $\ket{\psi_z^{(4)}}$,
labeled by $z\in\ZZ_2^4$, is orthogonal to
{its assigned stabilizer PVM ray}.}
\label{tab:sqrtT-four-zero-assignment}
\end{table}

{
\begin{proof}[Proof of Theorem~\ref{thm:recursive-cyclotomic-pbr}]
We recursively construct a measurement $M_k$.
For $k=3$, use the two-qubit ADM defined above.
For $k>3$, pair the $D_k$ qubits and proceed as follows.

\begin{enumerate}
\item Measure $ZZ$ on every pair, obtaining a syndrome
$\ell\in\{E,O\}^{D_k/2}$.

\item If $\ell$ contains an $O$, choose the first odd pair.
Measure logical $Y$ on this pair and logical $X$ on every
remaining pair.

\item If $\ell$ is all even, apply $M_{k-1}$ to the
$D_k/2=D_{k-1}$ encoded qubits, using
$\ket{0_E}=\ket{00}$ and $\ket{1_E}=\ket{11}$.
\end{enumerate}

To verify the WADM property, consider a complete outcome.
In step 2, the inputs $01$ and $10$ on the selected odd
pair give opposite logical $Y$ eigenstates.
Either logical-$Y$ outcome therefore excludes one of
these inputs, independently of the inputs and subsequent
outcomes on the remaining pairs.

In step 3, the physical inputs $00$ and $01$ give the
encoded states $\ket{\psi_0^{(k-1)}}$ and
$\ket{\psi_1^{(k-1)}}$.
By induction, every recursive outcome excludes a logical
preparation word. Replacing each logical $0$ by $00$
and each logical $1$ by $01$ gives an original preparation
word excluded by the complete outcome.

All logical measurements are physical Pauli measurements:
on each parity subspace, logical $X$, $Y$, and $Z$ can
be represented by $X_iX_j$, $Y_iX_j$, and $Z_i$,
respectively.
Each syndrome subspace is resolved into rank-one stabilizer
projectors, directly in step 2 and recursively in step 3.
Distinct syndrome subspaces are orthogonal, so the resulting
measurement is a rank-one stabilizer PVM.
Since every outcome excludes a preparation, it is a WADM.
\end{proof}
}

\subsection{Antidistinguishing measurements from complete graph states}

We now introduce a somewhat different construction that gives a stabilizer SADM for every $n\geq2$, with the preparation angle depending on $n$.
{This construction is closer to the original argument due to PBR}~\cite{pusey2012reality}.
In particular, we define
\[
\theta_n:=\frac{\pi}{4n},
\qquad
\ket{\psi_a}
:=
\cos\theta_n\ket0+(-1)^a\sin\theta_n\ket1,
\qquad a\in\ZZ_2.
\]
These states are nonorthogonal, since
$\langle\psi_0\lvert\psi_1\rangle=\cos(2\theta_n)>0$.
For $x\in\ZZ_2^n$, write
\[
\ket{\psi_x}:=\bigotimes_{j=1}^n\ket{\psi_{x_j}}.
\]

{We recall some facts about graph states;} see e.g.,~\cite{hein2006entanglement}.
{For a graph $G = (V,E)$ with vertex set $V = \{1,\cdots,n\}$ and edge set $E$, the corresponding graph state is defined by}
\[
\ket G = 2^{-n/2}\sum_{x\in\ZZ_{2}^{n}}(-1)^{\Gamma(x)}\ket x,
\quad\text{where}\quad
\Gamma(x) = \sum_{(i,j)\in E}x_{i}x_{j}.
\]
Every graph state is also a stabilizer state whose stabilizer group has the canonical generating set
\[
Q_{i}
:=
X_{i}\bigotimes_{j\in N(i)}Z_{j},
\quad
i=1,\cdots,n,
\]
where $N(i)\subset V$ is the neighborhood of the vertex $i$.
{The Pauli operator $Z_j$ commutes with $Q_i$ for $i\neq j$ and anticommutes with $Q_j$.}
Thus, we have the orthonormal stabilizer basis with rays
\[
Z^{x}\ket G,
\quad
Z^{x}
:=
\bigotimes_{j=1}^{n}Z^{x_{j}}.
\]
{The complete graph $K_n$ has an edge between every pair of distinct vertices.}
Its corresponding graph state is
\[
\ket{K_n}
=
2^{-n/2}\sum_{z\in\ZZ_2^n}
(-1)^{\binom{|z|}{2}}\ket z,
\]
since $\Gamma(z)=\sum_{i<j}z_i z_j=\binom{|z|}{2}$ and precisely $|z|$ entries of $z$ are equal to one.
We let the corresponding orthonormal basis rays be written as $\ket{\phi_{x}} := Z^{x}\ket{K_{n}}$.

\begin{thm}\label{thm:complete-graph-adm}
For every $n\geq2$, the product preparations
\[
\{\ket{\psi_{x}}\bra{\psi_{x}}:x\in\ZZ_2^n\}
\]
{admit a rank-one stabilizer ADM.}
\end{thm}

{%
\begin{proof}
For preparation $\ket{\psi_y}$, let
$d:=|x+y|$ be the Hamming distance between $x$ and $y$,
and write $c=\cos\theta_n$, $s=\sin\theta_n$.
Then
\[
\langle\phi_x\lvert\psi_y\rangle
=
2^{-n/2}
\sum_{z\in\ZZ_2^n}
(-1)^{\binom{|z|}{2}+(x+y)\cdot z}
c^{n-|z|}s^{|z|}.
\]
The signs $(-1)^{\binom w2}$ follow the repeating pattern
$1,1,-1,-1$, and hence
\[
(-1)^{\binom w2}
=\operatorname{Re}(i^w)+\operatorname{Im}(i^w).
\]
Applying this identity and factoring the sum over bits gives
\[
\langle\phi_x\lvert\psi_y\rangle
=
2^{-n/2}\left(
\operatorname{Re}\bigl[(c+is)^{n-d}(c-is)^d\bigr]
+
\operatorname{Im}\bigl[(c+is)^{n-d}(c-is)^d\bigr]
\right).
\]
Since $c\pm is=e^{\pm i\theta_n}$, this becomes
\[
\langle\phi_x\lvert\psi_y\rangle
=
2^{-n/2}
\left[
\cos\bigl((n-2d)\theta_n\bigr)
+
\sin\bigl((n-2d)\theta_n\bigr)
\right].
\]
For $\theta_n=\pi/(4n)$, the argument lies in
$[-\pi/4,\pi/4]$.
On this interval, $\cos t+\sin t$ vanishes only at
$t=-\pi/4$.
Consequently,
\[
\langle\phi_x\lvert\psi_y\rangle=0
\quad\Longleftrightarrow\quad
d=n
\quad\Longleftrightarrow\quad
y=x+\mathbf1.
\]
Thus each measurement outcome excludes exactly one
preparation, and each preparation has exactly one
excluding outcome.
In particular, the measurement is a SADM.
\end{proof}}

{Our construction yields a PBR proof for every number $n\geq2$ of qubits.}
Although less general than the SADM proposed by PBR, it uses
a measurement scheme built from Pauli measurements and graph
states.
The required angles $\theta_n=\pi/(4n)$ enter through the
preparations.
Whether this simplification of the measurement protocol makes actual implementation simpler requires separate consideration.

\section{Preparation independence in the Pauli ontological model}\label{sec:polytope}

We now specialize our preceding results to the Pauli ontological model~\cite{zurel2020hidden}, which furnishes a classical simulation algorithm for universal QCM based on convex polytopes.
(For preliminaries on convex polytopes, we refer the reader to~\cite{ziegler2012lectures}.)
Particularly relevant to any classical simulation algorithm is
its scaling in $n$, the number of qubits.
In many approaches, including those based on quasiprobability
distributions~\cite{pashayan2015estimating,howard2017application}
or stabilizer rank
methods~\cite{bravyi2016trading,bravyi2016improved,bravyi2019simulation},
tensor products provide a way to construct representations of
larger systems from those of their constituent parts.
{Here we use our WADMs for magic states to show that such tensor-product constructions in the Pauli ontological model encounter obstructions arising from the PBR argument.}

\subsection{The Pauli polytopes}

Let $S_n$ denote the set of pure $n$-qubit stabilizer states.
The $n$-qubit Pauli polytope, also called the $\Lambda$ polytope, is defined as the intersection of halfspace inequalities
\begin{eqnarray}
P_{n}
:=
\left\{ W\in \Herm(\hH) ~:~\Tr(W)=1,~\bra{\mathcal S}W\ket{\mathcal S}\geq 0~\forall~ \ket{\mathcal S}\in S_{n} \right \}.
\end{eqnarray}
This is a bounded polytope~\cite{zurel2020hidden}, and hence the convex
hull of its finitely many vertices~\cite{ziegler2012lectures}.
We label the extreme points $\Ext(P_{n})$ of $P_{n}$ by $\Lambda_{n}$.

Every quantum state belongs to $P_n$ and therefore admits a decomposition
\[
\rho
=
\sum_{\lambda\in\Lambda_n}
\mu_\rho(\lambda)A_\lambda,
\qquad
\mu_\rho(\lambda)\geq 0,
\qquad
\sum_{\lambda\in\Lambda_n}\mu_\rho(\lambda)=1.
\]
The labels $\lambda$ are interpreted as ontic states, and $\mu_\rho$ as a preparation distribution.
Such decompositions, however, are generally not unique.
A measurement of a Pauli observable $Q$ is represented by a stochastic
kernel $\Gamma$ satisfying
\[
\Pi_Q^\sigma A_\lambda\Pi_Q^\sigma
=
\sum_{\lambda'\in\Lambda_n}
\Gamma(\sigma,\lambda'\mid Q,\lambda)A_{\lambda'},
\]
where $\Pi_Q^\sigma$ is the projector associated with the outcome $\sigma\in\{\pm1\}$, and
\[
\Gamma(\sigma,\lambda'\mid Q,\lambda)\geq 0,
\qquad
\sum_{\sigma=\pm1}\sum_{\lambda'\in\Lambda_n}
\Gamma(\sigma,\lambda'\mid Q,\lambda)=1.
\]
The existence of these kernels follows from the behavior of $P_n$
under Pauli measurements; see~\cite{zurel2020hidden,okay2024classical}.

We are interested in whether independent preparations can be represented by independently sampled local ontic states. More technically, we study the corresponding operators combined by the tensor product.
To this end, one compatibility result has already been established. For any $A_\lambda\in\Ext(P_n)$, any stabilizer state $\ket{\mathcal S}\in S_m$, and any $(n+m)$-qubit Clifford unitary $U$,
the operator
\[
U\bigl(A_\lambda\otimes
\ket{\mathcal S}\bra{\mathcal S}\bigr)U^\dagger
\]
is a vertex of $P_{n+m}$~\cite{okay2021extremal}.
This construction adds only stabilizer ancillas and Clifford processing, and therefore introduces no additional computational resource.

{More relevant for quantum computation are tensor products of non-stabilizer resource states, such as those supplied by magic-state distillation protocols}~\cite{magic}.
For general preparations, consider decompositions
\[
\rho
=
\sum_{\lambda\in\Lambda_n}
\mu_\rho(\lambda)A_\lambda,
\quad \text{and} \quad
\rho'
=
\sum_{\lambda'\in\Lambda_m}
\mu_{\rho'}(\lambda')A_{\lambda'}.
\]
Their tensor product satisfies
\[
\rho\otimes\rho'
=
\sum_{\lambda\in\Lambda_n}
\sum_{\lambda'\in\Lambda_m}
\mu_\rho(\lambda)\mu_{\rho'}(\lambda')
\bigl(A_\lambda\otimes A_{\lambda'}\bigr).
\]
The coefficients form a product probability distribution, as required by preparation independence.
The remaining question is whether the tensored operators are valid elements of $P_{n+m}$.

Were we to have $A_{\lambda}\otimes A^{\prime}_{\lambda'}\in P_{n+m}$, then this construction yields an ontological model for universal QCM satisfying preparation independence.
Indeed, if $A_\lambda\otimes A_{\lambda'}\in P_{n+m}$, then every stabilizer
projective measurement $M$ has a normalized, non-negative response function
\[
\xi(k\lvert M,\lambda,\lambda')
=
\bra{\mathcal S^k}
A_\lambda\otimes A_{\lambda'}
\ket{\mathcal S^k},
\quad
\ket{\mathcal{S}_{k}}\in M.
\]
Moreover, if $A_\lambda\otimes A_{\lambda'}$ belongs to $P_{n+m}$, it is
in fact a vertex of $P_{n+m}$.%
\footnote{In the framework of
generalized probabilistic theories~\cite{plavala2023general},
the tensor product of two local extremal states is extremal
in the maximal tensor product
$P_n\hat\otimes P_m$; see also~\cite{okay2024classical}.
Since $P_{n+m}\subseteq P_n\hat\otimes P_m$, any such product
that belongs to $P_{n+m}$ is also extremal in $P_{n+m}$.
Thus membership implies extremality.}
Consequently, closure under tensor products of local vertices
would identify each pair $(\lambda,\lambda')$ with a global
ontic state and provide an operator realization of preparation
independence.

Our results up until this point suggest that preparation independence
should fail for the magic states considered thus far, provided that
their preparation distributions have overlapping support.
In the next sections, we show that the geometry of the Pauli polytopes
allows us to choose such distributions for distinct magic states.

\subsection{Convex geometry of Pauli polytopes}\label{sec:convex-structure}

The vertices that can appear in a decomposition of $W\in P_n$ are determined by its vanishing stabilizer probabilities.
{Indeed, if a stabilizer state $\ket{\mathcal S}$ satisfies $\bra{\mathcal S}W\ket{\mathcal S}=0$, then}
\[
\bra{\mathcal S}W\ket{\mathcal S}=0
=
\sum_{\lambda\in\Lambda_n}
\mu_W(\lambda)\bra{\mathcal S}A_{\lambda}\ket{\mathcal S}
=
0.
\]
{Since every term is nonnegative, this implies $\bra{\mathcal S}A_\lambda\ket{\mathcal S}=0$ for all $\lambda\in \Supp(\mu_W)$.}
These feasible vertices are elements of the \textit{carrier face} of $W$ introduced below; see also~\cite{rockafellar1997convex}.

Recall that $F\subset P_{n}$ is a \textit{face} of $P_{n}$ if it is of the form
\[
F_{V} = \{W\in P_{n}:~\bra{\mathcal S}W\ket{\mathcal S}=0,\quad \forall ~\ket{\mathcal S}\in V\},
\]
where $V\subset S_{n}$ is any subset of stabilizer states.
Note that $F$ can be the empty set.
{Fix some $W\in P_n$ and let $Z_W$ be the set of stabilizer rays with zero expectation on $W$.}
The carrier face of $W$ is then the inclusion minimal face of $P_{n}$ containing $W$.
It is given by
\[
F_W
:=
\left\{
A\in P_n:
\bra{\mathcal S}A\ket{\mathcal S}=0
\text{ for every }\ket{\mathcal S}\in Z_W
\right\}.
\]
In particular, we have that $W \in \operatorname{relint}(F_{W})$ is in the \textit{relative interior} of $F_{W}$~\cite{ziegler2012lectures}.
Geometrically, this means that there exists a ball of dimension $\text{dim}(F_{W})$ centered at $W$ with suitable radius $\varepsilon$ contained in $F_{W}$.

The vertices of $F_W$ are therefore precisely those vertices of $P_n$ satisfying
the above orthogonality conditions~\cite{rockafellar1997convex}.
We define the \emph{vertex support} of $W$ by
\[
\Lambda_W
:=
\{\lambda\in\Lambda_n:A_\lambda\in F_W\}.
\]
Every vertex decomposition of $W$ satisfies
\[
\Supp(\mu_W)\subseteq\Lambda_W.
\]
Conversely, every label in $\Lambda_W$ can occur with positive
probability in some decomposition of $W$.

\Lem{\label{lem:any-vertex}
Let $F$ be a nonempty face of $P_n$ and
$W\in\operatorname{relint}(F)$. For every
$A_\lambda\in\Ext(F)$, there exists a vertex decomposition of $W$
with $\mu_W(\lambda)>0$.
}

\begin{proof}
Since $W\in\operatorname{relint}(F)$, for sufficiently small
$\varepsilon\in(0,1)$ the point
\[
R=\frac{W-\varepsilon A_\lambda}{1-\varepsilon}
\]
belongs to $F$. Decomposing $R$ into vertices of $F$ and substituting
into
\[
W=\varepsilon A_\lambda+(1-\varepsilon)R
\]
gives the required decomposition.
\end{proof}

The following proposition identifies exactly when two operators
admit decompositions with a common ontic state.

\Pro{\label{pro:common-ontic-support}
For $W,W'\in P_n$, the following are equivalent:
\begin{enumerate}
\item There exist vertex decompositions $\mu_W,\mu_{W'}$ such that
\[
\Supp(\mu_W)\cap\Supp(\mu_{W'})\neq\varnothing.
\]
\item Their carrier faces satisfy
\[
F_W\cap F_{W'}\neq\varnothing.
\]
\end{enumerate}
In particular, if $F_W=F_{W'}$, then any vertex of this common face
can occur with positive coefficient in decompositions of both
$W$ and $W'$.
}

\begin{proof}
If $\lambda$ belongs to the support of both decompositions, then
$A_\lambda\in F_W\cap F_{W'}$.
Conversely, a nonempty intersection $F_W\cap F_{W'}$ is a face of $P_n$ and therefore contains a vertex $A_\lambda$ of $P_n$.
This is a vertex of both carrier faces.
Applying Lemma~\ref{lem:any-vertex} separately to $W$ and $W'$ one can construct preparation distributions such that $\mu_W(\lambda)>0$ and $\mu_{W'}(\lambda)>0$.
\end{proof}

{We can extend this result so that every vertex of $F_W$ appears simultaneously with strictly positive weight in a single decomposition.}

\begin{cor}\label{cor:carrier-spanning}
Let $F$ be a face of a polytope.
Every $W\in\operatorname{relint}(F)$ admits a decomposition
\[
W=\sum_{A\in\Ext(F)}p_A A,
\qquad
p_A>0,
\qquad
\sum_{A\in\Ext(F)}p_A=1.
\]
In particular, states with the same carrier face admit vertex decompositions with the same support.
\end{cor}

\begin{proof}
Let $\bar A$ be the uniform average of the vertices of $F$.
Since $W\in\operatorname{relint}(F)$, for sufficiently
small $\epsilon>0$ we have
\[
B:=\frac{W-\epsilon\bar A}{1-\epsilon}\in F.
\]
Expressing $B$ as a convex combination of the vertices of
$F$ and substituting into
$W=\epsilon\bar A+(1-\epsilon)B$
gives a strictly positive coefficient for every vertex.
\end{proof}

{Thus, to establish the existence of overlapping preparation distributions,} it suffices to show that the two states have the same carrier face.
Notice, however, that this is a statement about which admissible decompositions exist.
Not every pair of decompositions need overlap, even if the points share the same carrier face.
An exception is when the carrier face is a simplex since the extremal points are then affinely independent.

\subsection{Ontic supports for magic states}\label{sec:lambda-application}
{We now analyze the carrier faces of magic states,} with a particular emphasis on $T$-state preparations.
{A similar analysis for higher-level phases is straightforward.}
Introduce the notation
\[
\ket{T_n}:=\ket T^{\otimes n},
\qquad
\Pi_n:=\ket{T_n}\bra{T_n},
\qquad
\Pi_n^\perp:=Z^{\otimes n}\Pi_n Z^{\otimes n},
\qquad
F_n:=F_{\Pi_n}.
\]
Here $\Pi_n^\perp$ denotes the projector onto
$(Z\ket T)^{\otimes n}$, {rather than the complementary projector}
$I-\Pi_n$.
Our goal is to show that $\Pi_n$ and $\Pi_n^\perp$ have the same carrier face.
They will then admit vertex decompositions with common strictly positive support.

For a stabilizer ray $\ket\phi$, the Born probability
\[
\Tr(\Pi_n\ket\phi\bra\phi) =  \lvert \bra\phi T_{n} \rangle\rvert^{2}
\]
vanishes precisely when $\langle\phi\lvert T_n\rangle=0$.
{Recall the overlap in Eq.~\eqref{eq:cyclotomic-overlap} and specialize to}
$k=3$ and $z=0$.
Since $\zeta_3^2=i$, separate the even and odd Hamming weights of the overlap polynomial according to
\begin{eqnarray}
P_{\phi,0}(\zeta_3)=C_0+\zeta_3 C_1,
\qquad
C_r\in\QQ(i), \quad r=0,1.
\label{eq:t-carrier-parity}
\end{eqnarray}
Because $1$ and $\zeta_3$ are linearly independent over
$\QQ(i)$, a zero overlap requires $C_0=C_1=0$.
Physically, the amplitudes cancel separately in the even and
odd sectors.
Observe that applying the unitary $Z^{\otimes n}$ changes only the relative
sign of these sectors, so it cannot undo either cancellation.

\begin{pro}\label{pro:t-common-carrier}
For every $n\geq1$, the states $\Pi_n$ and $\Pi_n^\perp$
have the same carrier face $F_n$, and hence the same vertex support.
\end{pro}

\begin{proof}
The operator $Z^{\otimes n}$ multiplies the amplitude at $x$
by $(-1)^{|x|}$. The overlap with $Z^{\otimes n}\ket{T_n}$
is therefore proportional to $C_0-\zeta_3 C_1$.
By linear independence, this vanishes if and only if
$C_0=C_1=0$. Thus the two states have exactly the same
orthogonal stabilizer projectors and the same carrier face.
\end{proof}

The same argument permits a continuous change of relative
phase between the parity sectors. Let
\[
U_\alpha:=\exp\left(\frac{i\alpha}{2}Z^{\otimes n}\right)
=\cos\frac\alpha2\,I+i\sin\frac\alpha2\,Z^{\otimes n}.
\]

\begin{pro}\label{pro:carrier-deformation}
For every $n\geq1$ and $\alpha\in\mathbb R$,
$U_\alpha\Pi_nU_\alpha^\dagger\in F_n$.
If $e^{-i\alpha}\zeta_3\notin\QQ(i)$, the transformed state
has carrier face exactly $F_n$. In particular, this holds
for $\alpha=\pi$.
\end{pro}

\begin{proof}
The transformed overlap is proportional to
\begin{eqnarray}
e^{i\alpha/2}C_0+e^{-i\alpha/2}\zeta_3C_1
=e^{i\alpha/2}\left(C_0+e^{-i\alpha}\zeta_3C_1\right).
\label{eq:orthogonal-unitary}
\end{eqnarray}
Every original zero is preserved since $C_0=C_1=0$.
For $e^{-i\alpha}\zeta_3\notin\QQ(i)$, we have exact equality of the zero sets.
\end{proof}

{At exceptional angles, new stabilizer zeros can appear and}
the state can move to a proper subface.
{A second remark concerns symmetries.}
The Pauli polytopes are dual to the stabilizer polytopes and therefore have the extended Clifford group as their symmetry group~\cite{heinrich2019robustness}. The unitary above is therefore not a symmetry of $P_{n}$, except in special cases. Instead, it is a symmetry of a convex set within $F_{n}$.

\subsection{{Tensor obstruction for magic state supports}}
\label{sec:magic-support-tensors}

In this section we show that tensoring operator decompositions of $T$-states need not give a valid decomposition in the composite Pauli polytope.
The argument combines the carrier face results above with the two-block $T$-state protocol of Section~\ref{sec:partitioning-t-states}.
The underlying obstruction is that independent local sampling can produce a product operator that assigns a negative probability to a stabilizer measurement outcome.

Since $\Pi_m$ and $\Pi_m^\perp$ have the same carrier face, they admit decompositions with strictly positive weights on overlapping vertices~(Corollary~\ref{cor:carrier-spanning}).
Tensoring these decompositions with a fixed decomposition of $\Pi_n$ preserves this overlap.
However, the protocol of Section~\ref{sec:partitioning-t-states}, together with
Corollary~\ref{cor:magic-pi}, perfectly distinguishes these two composite preparations using adaptive Pauli measurements.
Consequently, they cannot share an ontic state with valid response functions.
At least one of the product operators in the common support must therefore lie outside the composite Pauli polytope.

\begin{pro}\label{pro:not-in-Pnn-v5}
For every $n,m\geq1$, there exist vertices
$A\in\Ext(F_n)$ and $B\in\Ext(F_m)$ such that
\[
A\otimes B\notin P_{n+m}.
\]
\end{pro}

\begin{proof}
Choose decompositions of $\Pi_n$ positive on every vertex
of $F_n$, and of both $\Pi_m$ and $\Pi_m^\perp$ positive
on every vertex of $F_m$.
The composite preparations
\[
\rho_0:=\Pi_n\otimes\Pi_m,
\qquad
\rho_1:=\Pi_n\otimes\Pi_m^\perp
\]
then have strictly positive decompositions over the product ontic space $\Ext(F_n)\times\Ext(F_m)$.

Suppose every $A\otimes B$ in this product set belongs to $P_{n+m}$.
Let $\{\Phi_l\}_l$ be the stabilizer measurement from
Corollary~\ref{cor:magic-pi}.
For each outcome $l$, at least one of $\Tr(\Phi_l\rho_0)$ and $\Tr(\Phi_l\rho_1)$ is zero.
Expanding that preparation gives a strictly positive sum of nonnegative terms $\Tr(\Phi_l(A\otimes B))$.
Every such term must vanish.
For any fixed pair of vertices, however,
\[
\sum_l\Tr(\Phi_l(A\otimes B))=\Tr(A\otimes B)=1,
\]
a contradiction.
\end{proof}

This argument shows only that some product operator violates a stabilizer inequality.
One could imagine, however, that a carefully chosen decomposition could avoid the problematic product operator(s), thereby salvaging the construction.
In the next section we show that this is not the case.

\subsection{{A general tensor obstruction for qubits}}
\label{sec:general-tensor-obstruction}

We now establish a stronger structural result about Pauli polytopes:
every tensor product of two qubit Pauli-polytope vertices lies outside the composite polytope.

For $A\in P_n$ and $B\in P_m$, every product stabilizer
state satisfies
\[
\bra{S_n}\bra{S_m}A\otimes B\ket{S_n}\ket{S_m}
=
\bra{S_n}A\ket{S_n}\bra{S_m}B\ket{S_m}
\geq 0.
\]
Thus, any violation must be witnessed by an entangled
stabilizer state.
For vertices, we construct such a witness by superposing
two carefully chosen product stabilizer states.
Each product state has a local factor with zero expectation
on the corresponding vertex.

\begin{lem}\label{lem:active-stabilizer-edge}
Let $A\in\Ext(P_n)$, with $n\geq1$.
There exist two orthogonal stabilizer states
$\ket{a_0},\ket{a_1}$, agreeing on $n-1$ independent
stabilizer generators and differing in the sign of one
remaining generator, such that
\[
\bra{a_0}A\ket{a_0}=0,
\qquad
\alpha:=\bra{a_0}A\ket{a_1}\neq0.
\]
Their phases can be chosen so that they are the encoded
computational basis of a rank-two stabilizer code.
\end{lem}

\begin{proof}
{Let $Z_A$ denote the set of \textit{tight} stabilizer rays, namely those satisfying $\bra a A\ket a=0$.}
Since $A$ is a vertex, the identity together with the projectors formed from $Z_{A}$ span the real vector space of Hermitian operators.
Otherwise, there would be a nonzero traceless Hermitian direction preserving all tight equalities, along which both sufficiently small positive and negative perturbations of $A$ remain in $P_n$, contradicting extremality; see e.g.,~\cite{chvatal1983linear}.

First, we need to show that not every $\ket{a}\in Z_{A}$ satisfies $A\ket{a} = 0$.
We proceed by contradiction.
Suppose instead that every tight stabilizer ray belongs to $K:=\ker A$.
{Since $\Tr A=1$, $A$ is not the zero operator and $\ker A$ is a proper subspace.}
This means we can choose unit vectors $\ket u\in K$ and $\ket v\in K^\perp$,
and define the Hermitian operator
\[
H:=\ket u\!\bra v+\ket v\!\bra u.
\]
{Observe that $\Tr H=0$ and $\bra aH\ket a=0$ for every tight}
stabilizer ray $\ket a$, since $\ket a\in K$.
In particular, we have $\bra a (A\pm\varepsilon H) \ket a = 0$ for all $\ket a \in Z_{A}$ so that all tight equalities are preserved along $H$.
Since all remaining stabilizer inequalities are strict, there exists a sufficiently small $\varepsilon > 0$ such that $A\pm\varepsilon H\in P_n$.
This contradicts extremality.
Hence some tight stabilizer ray $\ket{a_0}$ satisfies
$A\ket{a_0}\neq0$.

Recall that any stabilizer state is of the form $\ket{a_0} = U \ket 0^{\otimes n}$, for some Clifford unitary $U$.
Complete $\ket{a_{0}}$ to an orthonormal stabilizer basis $\mathcal B$ by applying $U$ to the computational basis.
Since $\ket{a_{0}}$ is not in the kernel of $A$ and because $\mathcal B$ is a basis, there exists some $\ket{a_{1}}\in \mathcal B$ such that $\bra{a_0}A\ket{a_1}\neq0$.
It is known that two stabilizer rays related by the same unitary to the computational basis can rearrange their stabilizer generators so that they differ in the sign of just a single generator. Removing that generator yields a rank-two stabilizer code.
\end{proof}

\begin{thm}\label{thm:qubit-vertex-tensor}
For all $n,m\geq1$ and all vertices
$A\in\Ext(P_n)$ and $B\in\Ext(P_m)$,
\[
A\otimes B\notin P_{n+m}.
\]
More precisely, there is an entangled rank-one stabilizer
projector $\Pi$ with $\Tr(\Pi(A\otimes B))<0$.
\end{thm}

\begin{proof}
Apply Lemma~\ref{lem:active-stabilizer-edge} to $A$ and $B$.
Choose encoded stabilizer pairs satisfying
\[
\bra{a_0}A\ket{a_0}=0,
\quad \alpha:=\bra{a_0}A\ket{a_1}\neq0,
\qquad
\bra{b_0}B\ket{b_0}=0,
\quad \beta:=\bra{b_0}B\ket{b_1}\neq0.
\]
Put $W:=A\otimes B$ and consider the product states
\[
\ket u:=\ket{a_0}\otimes\ket{b_1},
\qquad
\ket v:=\ket{a_1}\otimes\ket{b_0}.
\]
Their expectations vanish:
\[
\bra uW\ket u=\bra vW\ket v=0.
\]
Their off-diagonal matrix element, however, is
\[
c:=\bra uW\ket v=\alpha\overline\beta\neq0.
\]

Let $Z_A$ and $Z_B$ denote the logical Pauli operators
distinguishing the two states in each local pair.
The vectors $u,v$ share the $n+m-2$ local code generators
and the additional generator $-Z_AZ_B$.
They therefore span a rank-two stabilizer code.
With the phases fixed by the local Clifford encodings,
its complementary logical Pauli bases consist of
\[
\ket{x_\pm}:=\frac{\ket u\pm\ket v}{\sqrt2},
\qquad
\ket{y_\pm}:=\frac{\ket u\pm i\ket v}{\sqrt2}.
\]
These are rank-one stabilizer states, each entangled across
the $n|m$ partition. Direct calculation gives
\[
\bra{x_\pm}W\ket{x_\pm}=\pm\text{Re}(c),
\qquad
\bra{y_\pm}W\ket{y_\pm}=\mp\text{Im}(c).
\]
Since $c\neq0$, at least one of these expectations is
strictly negative. In fact, the violating projector can
be chosen so that
\[
\Tr(\Pi W)
=-\max\{|\text{Re}(\alpha\overline\beta)|,
|\text{Im}(\alpha\overline\beta)|\}<0.
\]
Thus $A\otimes B$ violates a defining stabilizer inequality
of $P_{n+m}$.
\end{proof}

Consequently, independently sampling local vertex decompositions and returning the operator $A\otimes B$ \textit{never} produces a valid composite ontic operator.
In short: preparation independence fails for the Pauli ontological model.

\section{{Conclusion}}\label{sec:conclusion}

In this paper we initiated the development of PBR proofs for non-stabilizer preparation families using Pauli measurements.
For the single-qubit preparation families considered at levels $k\geq3$ of the Clifford hierarchy, we showed that stabilizer-restricted PBR proofs exist using $D_k=2^{k-2}$ independently prepared systems.
Beyond these Clifford-hierarchy families, we also constructed such measurement schemes for every $n\geq2$, with the preparation states depending on $n$.
Although these constructions do not necessarily provide new conceptual insights into the PBR theorem, they open an avenue for studying particularly simple and elegant proofs, similar in spirit to the study of Kochen--Specker proofs~\cite{cabello1996bell,waegell2012proofs}.
Particularly interesting is the role of algebraic number theory
in the geometry of stabilizer-restricted antidistinguishability,
which our work has only begun to explore.
Related connections between number theory and quantum-state geometry
have been developed, for instance, in the study of SIC measurements~\cite{appleby2017sics}.

We also demonstrated a general obstruction to preparation independence
in ontological representations of qubit stabilizer theory.
Our results imply that any such model satisfying preparation independence
must be $\psi$-ontic on pure stabilizer states, suggesting the stabilizer
states themselves as a natural candidate ontic state space.
Indeed, these states also serve as the underlying state space in
quasiprobability representations of qubit stabilizer
theory~\cite{howard2017application}.
This contrasts with the odd-prime-dimensional qudit case, where
stabilizer states have nonnegative Gross Wigner representations
and the phase-point operators are closed under tensor products.
Several distinctions between the even- and odd-dimensional cases
have been related to topological obstructions~\cite{raussendorf2023role}.
Whether related obstructions also account for the difference in
tensor-product structure appears plausible (see Appendix~\ref{sec:contextuality-pbr}), but remains an interesting question for future work.

Our results also connect the operational constraints imposed by
antidistinguishability to the geometry of Pauli polytopes.
In this representation, the obstruction to preparation independence
appears concretely as a failure of closure under tensor products:
the tensor product of any two local vertices lies outside the
corresponding composite-system polytope.
This restricts how these polytopes can represent independently
prepared systems.
From a categorical perspective, it appears to obstruct a monoidal representation
in which parallel
composition is implemented by the ordinary operator tensor product.
While a full categorical treatment of polyhedral classical simulation schemes~\cite{okay2026polyhedral} remains open, some steps in this direction were taken in~\cite{okay2025double}.

This question also bears on the computational complexity of polytope-based classical simulation.
Zurel et al.~\cite{zurel2024simulating} showed that the simulation
requires only polynomially many bits to describe its instantaneous
state, leaving the presumed computational hardness in the initial
sampling or subsequent measurement updates.
An efficiently implementable tensor-product rule would make
initialization of fixed single-qubit product inputs straightforward:
one could sample the local distributions independently and compose
the resulting ontic states.
Our result rules out this construction using operator tensor products
of local vertices.
It does not establish the hardness of initial sampling, but it does
remove a natural reason to expect initialization to be easy.
The difficulty of classical simulation may therefore arise before
the first measurement, in representing completely unentangled inputs
within an ontological state space not admitting a product structure.

\paragraph*{Acknowledgments}
The author thanks P.~Mitosek, R.~Raussendorf, and A.~T.~Y\"ucel
for useful discussions.
The author acknowledges support from the Nieders\"achsisches
Ministerium f\"ur Wissenschaft und Kultur and the Alexander von
Humboldt Foundation.
The author acknowledges the use of ChatGPT in literature review,
manuscript revision, and the development of mathematical arguments.
The author has verified the final results and takes full responsibility
for the manuscript.

\bibliography{bib2}
\bibliographystyle{ieeetr}

\appendix

\section{Algebra background}\label{sec:field-ext}

Here we recall the facts about field extensions and cyclotomic fields
used in the main text. For a detailed treatment,
see, e.g.,~\cite{herstein1991topics}.

\subsection{Field extensions}

Let $\FF$ be a field and let $\alpha$ be algebraic over $\FF$,
meaning that it is a root of a nonzero polynomial in $\FF[x]$.
The minimal polynomial $m_\alpha(x)$ is the unique monic polynomial
of least degree satisfying $m_\alpha(\alpha)=0$.
It is irreducible over $\FF$, and
\[
P(\alpha)=0
\quad\Longleftrightarrow\quad
m_\alpha(x)\mid P(x)
\qquad\text{for }P\in\FF[x].
\]
In particular, no nonzero polynomial of degree less than
$\deg m_\alpha$ can vanish at $\alpha$.

The field $\FF(\alpha)$ is the smallest field containing
$\FF$ and $\alpha$. Its degree over $\FF$ is its dimension
as an $\FF$-vector space, and
\[
[\FF(\alpha):\FF]=\deg m_\alpha.
\]
For a tower of finite extensions $\FF\subseteq K\subseteq L$,
the degrees satisfy
\[
[L:\FF]=[L:K]\,[K:\FF].
\]

\subsection{Cyclotomic fields}

A primitive $r$th root of unity is $\xi_r=e^{2\pi i/r}$.
Its minimal polynomial over $\QQ$ is the cyclotomic polynomial
\[
\Phi_r(x)
=
\prod_{\substack{1\leq j\leq r\\ \gcd(j,r)=1}}
(x-\xi_r^j).
\]
This polynomial is irreducible over $\QQ$ and has degree
$\varphi(r)$, where $\varphi$ is Euler's totient function.
Consequently,
\[
[\QQ(\xi_r):\QQ]=\varphi(r).
\]

In the main text we use $\zeta_k=\xi_{2^k}$.
For $k\geq2$, we have $i=\zeta_k^{2^{k-2}}$, so
$\QQ(i)\subseteq\QQ(\zeta_k)$.
The tower law therefore gives
\[
[\QQ(\zeta_k):\QQ(i)]
=
\frac{\varphi(2^k)}{[\QQ(i):\QQ]}
=
2^{k-2}
=:D_k.
\]
Since $\zeta_k^{D_k}=i$, the polynomial $x^{D_k}-i$
vanishes at $\zeta_k$ and has precisely this degree.
It is therefore the minimal polynomial of $\zeta_k$
over $\QQ(i)$.
Thus, for $P\in\QQ(i)[x]$,
\[
P(\zeta_k)=0
\quad\Longleftrightarrow\quad
x^{D_k}-i\mid P(x).
\]
In particular, if $\deg P<D_k$, then $P(\zeta_k)=0$
only when $P$ is the zero polynomial.
This is the algebraic constraint used in the
stabilizer-overlap argument.

\section{Contextuality-based PBR theorem for stabilizer states}%
\label{sec:contextuality-pbr}

We now give a contextuality-based proof using the notation and conventions of the main body together with the results of Karanjai et al.~\cite{karanjai2018contextuality}, which connect contextuality to partitioning measurements.
For a preparation family $P=\{\rho_s:s\in S\}$, let $\Omega_P$ be the set of Hermitian Pauli observables whose measurement has a deterministic outcome on at least one
$\rho_s$.
Denote its closure under products of commuting observables by $\widetilde\Omega_P$.
The relevant Kochen--Specker relations are
\[
\nu(QR)=\nu(Q)\nu(R)
\quad\text{if }[Q,R]=0,
\qquad
\nu(\pm I)=\pm1.
\]
{These relations express the algebraic constraints on compatible measurement outcomes.}

In the ontological model, preparations $\rho_s$ are represented
by probability distributions $\mu_s$.
We assume that the response functions are noncontextual under
fine-graining: if $N=\{\Pi_N^j\}_j$ refines
$M=\{\Pi_M^k\}_k$, with
\[
\Pi_M^k=\sum_{j:f(j)=k}\Pi_N^j,
\]
then
\[
\xi(k\lvert M,\lambda)
=
\sum_{j:f(j)=k}\xi(j\lvert N,\lambda).
\]
In particular, the response to a coarse-grained outcome is
independent of the chosen refinement.

Suppose now that there exists some $\lambda$ in the joint support of the preparations, i.e.
$\lambda\in\bigcap_{s\in S}\Supp(\mu_s)$.
For every $Q\in\Omega_P$, some preparation has a deterministic
outcome $\nu(Q)\in\{\pm1\}$.
Nonnegativity of the ontological representation therefore gives
\[
\xi(k\lvert M_Q,\lambda)=\delta_{k,\nu(Q)}.
\]
If two preparations impose different values on the same
observable, this already excludes a common $\lambda$.
Now let $Q,R$ be commuting observables whose responses at
$\lambda$ are deterministic.
Their joint measurement refines both $M_Q$ and $M_R$, so its response is concentrated on $(\nu(Q),\nu(R))$.
The same joint measurement refines $M_{QR}$, with product outcome $qr$.
Consequently,
\[
\xi(k\lvert M_{QR},\lambda)
=
\delta_{k,\nu(Q)\nu(R)}.
\]
Iterating this argument extends the deterministic assignment to $\widetilde\Omega_P$, where it satisfies
\[
\nu(QR)=\nu(Q)\nu(R)
\quad\text{whenever }[Q,R]=0,
\qquad
\nu(\pm I)=\pm1.
\]
If $\widetilde\Omega_P$ admits a proof of state-independent contextuality, no such assignment exists.
Thus we conclude
\[
\bigcap_{s\in S}\Supp(\mu_s)=\emptyset.
\]
Similar notions have been developed in~\cite{karanjai2018contextuality}.

Consider now two distinct, nonorthogonal pure stabilizer preparations $\rho_0,\rho_1$, with stabilizer groups $\mathcal S_0,\mathcal S_1$.
There exist anticommuting $g\in\mathcal S_0$ and $h\in\mathcal S_1$.
Otherwise, maximality would imply that the two groups have the same unsigned observables, making the states either identical or orthogonal.
Define the preparations
\[
\rho_{ab}:=\rho_a\otimes\rho_b,
\qquad
P:=\{\rho_{ab}:a,b\in\ZZ_2\}.
\]
{Since these are stabilizer states, the set of observables $\Omega_P$ and its closure $\widetilde\Omega_P$ are nonempty.}

\begin{pro}\label{pro:non-orthogonal-state-ind}
The closure $\widetilde\Omega_P$ contains a Mermin square.
Consequently, every ontological model reproducing the
allowed measurements and state updates satisfies
\[
\bigcap_{a,b\in\ZZ_2}\Supp(\mu_{\rho_{ab}})=\emptyset.
\]
\end{pro}

\begin{proof}
Put $l:=igh$, so that $l$ is a Hermitian Pauli observable.
Consider the array
\[
\begin{array}{c|c|c}
g\otimes I&I\otimes g&g\otimes g\\ \hline
I\otimes h&h\otimes I&h\otimes h\\ \hline
g\otimes h&h\otimes g&l\otimes l
\end{array}.
\]
This is precisely the structure of the Peres-Mermin square~\cite{mermin1993hidden}.
The first eight observables belong to $\Omega_P$, each
having deterministic value $+1$ on some preparation.
The final observable $L:=l\otimes l$ belongs to
$\widetilde\Omega_P$ since
\[
(g\otimes h)(h\otimes g)=L,
\qquad
(g\otimes g)(h\otimes h)=-L.
\]
Each row and column consists of commuting observables.
The row products are $I$, while the column products are
$I,I,-I$, precluding a consistent value assignment.
By the construction of Karanjai et al.
\cite[Theorem~2]{karanjai2018contextuality}, $M_L$ is a
partitioning measurement, thus excluding common ontic support for the states in $P$.
\end{proof}

\begin{cor}\label{thm:stabilizer-prep-ind}
An ontological model reproducing qubit stabilizer theory
and satisfying preparation independence must represent
distinct pure stabilizer preparations $\rho_0,\rho_1$
by distributions satisfying
\[
\Supp(\mu_{\rho_0})\cap\Supp(\mu_{\rho_1})=\emptyset.
\]
\end{cor}

\begin{proof}
Orthogonal stabilizer preparations are SSD and therefore
have disjoint supports.
For a distinct nonorthogonal pair, the result follows from Proposition~\ref{pro:non-orthogonal-state-ind}.
\end{proof}

Thus a model with overlapping distributions for distinct
pure qubit stabilizer states must violate preparation
independence.
Note that a similar result for qudit stabilizer theory with odd local dimension does not hold.
In that case the Wigner representation yields a noncontextual model that is also closed under tensor product, thus satisfying preparation independence.
As a concluding remark, we note that the Mermin-Peres square paradox~\cite{mermin1993hidden} can be understood topologically~\cite{okay2017topological}.

\end{document}